\documentclass
[preprint,10pt,twocolumn,notitlepage,showpacs,nofootinbib,rmp,groupedaddress]{revtex4}%
\usepackage{amsfonts}
\usepackage{amsmath}
\usepackage{amssymb}
\usepackage{graphicx}%
\PassOptionsToPackage{linktocpage}{hyperref}

\begin{document}
\title{Basis of a non Riemannian Geometry within the Equilibrium Thermodynamics}
\author{L. Velazquez}
\affiliation{Departamento de F\'{\i}sica, Universidad de Pinar del Rio, Marti 270, Esq. 27
de Noviembre, Pinar del Rio, Cuba}
\author{F. Guzman}
\affiliation{Instituto Superior de Tecnologia y Ciencias Aplicadas, Quinta de los Molinos,
Plaza de la Revolucion, La Habana, Cuba}

\begin{abstract}
Microcanonical description is characterized by the presence of an internal
symmetry closely related with the dynamical origin of this ensemble:
\textit{the reparametrization invariance}. Such symmetry possibilities the
development of a non Riemannian geometric formulation within the
microcanonical description of an isolated system, which leads to an unexpected
generalization of the Gibbs canonical ensemble and the classical fluctuation
theory for the open systems, the improvement of Monte Carlo simulations based
on the canonical ensemble, as well as a reconsideration of any classification
scheme of the phase transitions based on the concavity of the microcanonical entropy.

\end{abstract}
\date{\today}
\maketitle
\tableofcontents

\section{Introduction}

Generally speaking, a phase transition brings about a sudden change of the
macroscopic properties of a system while smoothly varying of a control
parameter. The mathematical description of phase transitions in the
conventional Thermodynamics is based on the \textit{loss of analyticity} (the
appearance of a singularity in the firsts derivatives) of the (gran)canonical
thermodynamic potential \cite{gallavotti,Gold}, which are related with the
existence of zeros in the partition function appearing with the imposition of
the thermodynamic limit \cite{yang-lee}.

However, phase transitions are also relevant in systems outside the
thermodynamic limit. A nontrivial example is the \textit{nuclear
multifragmentation}, a phenomenon taking place as a consequence of peripheral
collisions of heavy ions, which can be classified under the light of the new
developments as a first-order phase transition \cite{moretto,Dagostino}.

The key for taking into account the phase transitions in isolated systems
outside the thermodynamic limit is the consideration of the well-known
\textit{Boltzmann Principle}:%
\begin{equation}%
\begin{tabular}
[c]{|c|}\hline
$S_{B}=\ln W$\\\hline
\end{tabular}
\ \ \ \ \ \ \label{SB}%
\end{equation}
his celebrated gravestone epitaph. Since $W$ is the number of microscopic
states compatible with a given macroscopic state, the Boltzmann entropy
\ref{SB} is a measure of the size of the microcanonical ensemble. Within
Classical Statistical Mechanics $W$ is just the microcanonical accessible
phase space volume, that is, a geometric quantity. This explains why the
entropy \ref{SB} does not satisfy the concavity and the extensivity
properties, neither demands the imposition of the thermodynamic limit or a
probabilistic interpretation like the Shannon-Boltzmann Gibbs extensive
entropy:%
\begin{equation}
S_{e}=-\sum_{k}p_{k}\ln p_{k}. \label{sbg}%
\end{equation}
As already shown by Gross \cite{gro1,gro na}, phase transitions in small
systems could be classified within the microcanonical ensemble throughout the
"topology" of the \textit{Hessian} of the Boltzmann entropy.

We will show in the present work that the microcanonical description is
characterized by the existence of an internal symmetry: \textit{the
reparametrization invariance}. We shall demonstrate that the presence of this
symmetry implies a revision of classification of phase transitions based on
the concavity of the entropy and the theory of the statistical ensembles
derived from microcanonical basis.

\section{Reparametrization invariance\label{RI}}

\subsection{Geometrical basis}

Universality of the microscopic mechanisms of chaoticity provides a general
background for justifying the necessary ergodicity which supports a
thermostatistical description with microcanonical basis for all those
nonintegrable many-body Hamiltonian systems \cite{PF1,PF2,lieberman,pettini
51,cohenG}. Thus, the microcanonical ensemble:%
\begin{equation}
\hat{\omega}_{M}\left(  I,N\right)  =\frac{1}{\Omega\left(  I,N\right)
}\delta\left\langle I-\hat{I}\left(  X\right)  \right\rangle , \label{micro}%
\end{equation}
is just a \textit{dynamical ensemble} where every macroscopic characterization
has a direct mechanical interpretation. Here, $X$ represents a given point of
the phase space $\Gamma$ and $\hat{I}\left(  X\right)  =\left\{  \hat{I}%
^{1}\left(  X\right)  ,\hat{I}^{2}\left(  X\right)  ,\ldots\hat{I}^{n}\left(
X\right)  \right\}  $ are all those relevant (analytical) integrals of motion
determining the microcanonical description in a given application (since
Poincare-Fermi theorem \cite{PF1,PF2}: the\ total energy, the angular and
linear momentum).

The admissible values of the set of integrals of motion $\hat{I}\left(
X\right)  $ could be considered as the "coordinate points" $I=\left\{
I^{1},I^{2},\ldots I^{n}\right\}  $\ of certain subset\ $\mathcal{R}_{I}$\ of
the n-dimensional Euclidean space $\mathcal{R}^{n}$. Each of these points
determines certain sub-manifold $\mathcal{S}_{p}$ of the phase space $\Gamma$
:%
\begin{equation}
X\in\mathcal{S}_{p}\equiv\left\{  X\in\Gamma\left\vert \forall k~I^{k}\left(
X\right)  =I^{k}\right.  \right\}  , \label{set}%
\end{equation}
in which the system trajectories spread uniformly in accordance with the
ergodic character of the microscopic dynamics. Such sub-manifolds defines a
partition $\Im$ of the phase space $\Gamma$ in disjoint sub-manifolds:
\begin{equation}
\Im=\left\{  \mathcal{S}_{p}\subset\Gamma\left\vert ~%
{\displaystyle\bigcup_{p}}
\mathcal{S}_{p}=\Gamma~;~\mathcal{S}_{p}\cap\mathcal{S}_{q}=\varnothing
\right.  \right\}  . \label{partition}%
\end{equation}
Definitions \ref{set} and \ref{partition} allow the existence of a bijective
map $\psi_{I}$ between the elements of $\Im$ (sub-manifolds $\mathcal{S}%
_{p}\subset\Gamma$) and the elements of $\mathcal{R}_{I}$ (points
$I\in\mathcal{R}^{n}$):%

\begin{equation}
\psi_{I}:\Im\rightarrow\mathcal{R}_{I}\equiv\left\{  \forall\mathcal{S}_{p}%
\in\Im\left(  \Gamma\right)  ~\exists I\in\mathcal{R}_{I}\subset
\mathcal{R}^{n}\right\}  .
\end{equation}
Thus, the partition $\Im$ has the same topological features of the
n-dimensional Euclidean subset $\mathcal{R}_{I}$. For this reason $\Im$ will
be referred as the \textit{abstract space of the integrals of motions}. We say
that the map $\psi_{I}$ defines the n-dimensional Euclidean \textit{coordinate
representation} $\mathcal{R}_{I}$ of the\ abstract space $\Im$.

Let us now to consider another subset $\mathcal{R}_{\varphi}\subset
\mathcal{R}^{n}$ with the same \textit{diffeomorphic structure} of the subset
$\mathcal{R}_{I}$ and the following diffeormorphic map $\varphi$ among them:%
\begin{equation}
\varphi:\mathcal{R}_{I}\rightarrow\mathcal{R}_{\varphi}\equiv\left\{  \forall
I\in\mathcal{R}_{I}~\exists\varphi\in\mathcal{R}_{\varphi}\left\vert
\det\left(  \frac{\partial\varphi^{j}}{\partial I^{k}}\right)  \not =0\right.
\right\}  . \label{diffeomorphic}%
\end{equation}
We say that the map $\varphi$ represents a general \textit{reparametrization
change} of the microcanonical description since it allows us to introduce
another n-dimensional Euclidean coordinate representation $\mathcal{R}%
_{\varphi}$ by considering the bijective map $\psi_{\varphi}=\psi_{I}%
o\varphi^{-1}$:%
\begin{equation}
\psi_{\varphi}:\Im\rightarrow\mathcal{R}_{\varphi}\equiv\left\{
\forall\mathcal{S}_{p}\in\Im~\exists\varphi\in\mathcal{R}_{\varphi}%
\subset\mathcal{R}^{n}\right\}  .
\end{equation}
The above reparametrization change $\varphi$ also induces the following
reparametrization of the relevant integrals of motion $\varphi_{X}:\hat
{I}\left(  X\right)  \rightarrow\hat{\varphi}\left(  X\right)  $, where:
\begin{equation}
\hat{\varphi}\left(  X\right)  =\left\{  \varphi^{1}\left\langle \hat
{I}\left(  X\right)  \right\rangle ,\varphi^{2}\left\langle \hat{I}\left(
X\right)  \right\rangle ,\ldots\varphi^{n}\left\langle \hat{I}\left(
X\right)  \right\rangle \right\}  .
\end{equation}
Since $\hat{I}\left(  X\right)  $ are integrals of motions, every $\varphi
^{k}\left\langle \hat{I}\left(  X\right)  \right\rangle \in\hat{\varphi
}\left(  X\right)  $ will be also an integral of motion. The bijective
character of the reparametrization change $\varphi:\mathcal{R}_{I}%
\rightarrow\mathcal{R}_{\varphi}$ allows us to say that the sets $\hat
{\varphi}\left(  X\right)  $ and $\hat{I}\left(  X\right)  $ are
\textit{equivalent representations} of the relevant integrals of motion of the
microcanonical description because of they generate the same phase space
partition $\Im$.

The interesting question is that the microcanonical ensemble is
\textit{invariant} under every reparametrization change. Considering the
identity:%
\begin{equation}
\delta\left\langle \varphi-\hat{\varphi}\left(  X\right)  \right\rangle
\equiv\left\vert \frac{\partial\varphi}{\partial I}\right\vert ^{-1}%
\delta\left\langle I-\hat{I}\left(  X\right)  \right\rangle ,
\end{equation}
where $\left\vert \partial\varphi/\partial I\right\vert \not =0$ is the
Jacobian of the reparametrization change $\varphi$, the phase space
integration leads to the following transformation rule for the microcanonical
partition function:%
\begin{equation}
\Omega\left(  \varphi,N\right)  =\left\vert \frac{\partial\varphi}{\partial
I}\right\vert ^{-1}\Omega\left(  I,N\right)  ,
\end{equation}
leading in this way to the reparametrization invariance of the microcanonical
distribution function:%
\begin{equation}
\frac{1}{\Omega\left(  \varphi,N\right)  }\delta\left\langle \varphi
-\hat{\varphi}\left(  X\right)  \right\rangle \equiv\frac{1}{\Omega\left(
I,N\right)  }\delta\left\langle I-\hat{I}\left(  X\right)  \right\rangle .
\label{invariance}%
\end{equation}

A corollary of the identity \ref{invariance} is that the Physics derived from
the microcanonical description is reparametrization invariant since the
expectation values $\left\langle O\right\rangle $ of any macroscopic
observable $\hat{O}\left(  X\right)  $ obtained from the microcanonical
distribution function $\hat{\omega}_{M}\left(  X\right)  $ exhibits this kind
of symmetry:%
\begin{equation}
\left\langle O\right\rangle =\int\hat{O}\left(  X\right)  \hat{\omega}%
_{M}\left(  X\right)  dX\Rightarrow\left\langle O\right\rangle \left(
\varphi,N\right)  =\left\langle O\right\rangle \left(  I,N\right)  .
\end{equation}
The reparametrization invariance does not introduce anything new in the
macroscopic description of a given system, except the possibility of
describing the microcanonical macroscopic state by using any coordinate
representation of the abstract space $\Im$, a situation analogue to the
possibility of describing the physical space $\mathcal{R}^{3}$ by using a
Cartesian coordinates $\left(  x,y,z\right)  $ or a spherical coordinates
$\left(  r,\theta,\varphi\right)  $. Thus, we can develop a geometrical
formulation of Thermostatistics within the microcanonical ensemble.

\subsection{Covariant transformation rules}

The microcanonical partition function allows us to introduce an invariant
measure $d\mu=\Omega dI$ for the abstract space $\Im$, leading in this way to
an invariant definition of the Boltzmann entropy $S_{B}=\ln W$, where
$W=\int_{\Sigma_{\alpha}}d\mu$ \ characterizes certain \textit{coarse grained}
partition $\left\{  \Sigma_{\alpha}\left\vert ~%
{\textstyle\bigcup\nolimits_{\alpha}}
\Sigma_{\alpha}=\Im\right.  \right\}  $. In the thermodynamic limit
$N\rightarrow\infty$ the coarsed grained nature of the Boltzmann entropy can
be disregarded and taken as \textit{a scalar function} defined on the space
$\Im$ (see the appendix \ref{ent})

Since the Boltzmann entropy is a scalar function, its first derivatives obey
the transformation rule of a covariant vector during the reparametrization
changes:%
\begin{equation}
\frac{\partial S_{B}}{\partial\varphi^{K}}=\frac{\partial I^{S}}%
{\partial\varphi^{K}}\frac{\partial S_{B}}{\partial I^{S}}.
\label{covariant_vector}%
\end{equation}
However, the second derivatives (Hessian) of the entropy do not correspond to
a second rank covariant tensor:%
\begin{equation}
\frac{\partial^{2}S_{B}}{\partial\varphi^{K}\partial\varphi^{L}}%
=\frac{\partial I^{S}}{\partial\varphi^{K}}\frac{\partial I^{T}}%
{\partial\varphi^{L}}\frac{\partial^{2}S_{B}}{\partial I^{K}\partial I^{T}%
}+\frac{\partial^{2}I^{S}}{\partial\varphi^{K}\partial\varphi^{L}}%
\frac{\partial S_{B}}{\partial I^{S}}, \label{hessian_tr}%
\end{equation}
because of the correct transformation rule should be given by:%
\begin{equation}
\tau_{KL}^{\prime}=\frac{\partial I^{S}}{\partial\varphi^{K}}\frac{\partial
I^{T}}{\partial\varphi^{L}}\tau_{ST}.
\end{equation}

Covariant tensors can be derived from the differentiation of vectors within a
Riemannian geometry throughout the introduction of the covariant
differentiation, which depends on the existence of an appropriate metric.
There are in the past other geometric formulations of the thermodynamics which
identify the metric with Hessian of entropy. We provide in the appendix
\ref{ruppeiner} a little explanation of the different underlying physics
supporting such geometrical constructions. A complete review of such
formulations can be seen in ref.\cite{rupper}.

The using of the Hessian as a metric will be discarded in the present approach
due to the Hessian of the entropy associated to an isolated Hamiltonian system
does not correspond to a second rank covariant vector. Reader may think that
the non reparametrization invariance of the entropy Hessian is a ugly defect
of the present geometric formulation. Contrary, we will show that this feature
could become a big \textit{advantage} in understanding the nature of the phase
transitions within the microcanonical ensemble, as well as in the improving of
some Monte Carlo methods based on the Statistical Mechanics.

With some important exceptions like the astrophysical systems, the nuclear,
atomic and molecular clusters, and some other systems, most applications of
statistical mechanics are concerned with the behavior of atoms and molecules
subjected to short-range forces of electrostatic origin, i.e. gases, liquids,
and solids. These systems are typically enclosed by rigid boundaries, and
consequently, the only relevant integral of motion here is the total energy
$E$. It is very easy to verify that any bijective application of the total
energy $\Theta=\Theta\left(  E\right)  $ ensures the reparametrization
invariance of the microcanonical description:
\begin{equation}
\frac{1}{\Omega\left(  \Theta,N\right)  }\delta\left\langle \Theta-\hat
{\Theta}_{N}\right\rangle =\frac{1}{\Omega\left(  E,N\right)  }\delta
\left\langle E-\hat{H}_{N}\right\rangle ,
\end{equation}
being $\hat{\Theta}_{N}\equiv\Theta\left(  \hat{H}_{N}\right)  $, and
consequently, the results obtained above are still applicable in this context.

Summarizing the result of the present section: (1) The microcanonical
description is reparametrization invariant; (2) The Boltzmann entropy is just
a scalar function under the reparametrizations; (3) While the first
derivatives of the entropy are the components of a covariant vector, the
second derivatives do not correspond to a second rank covariant tensor. The
Boltzmann entropy also allows the introduction of other thermodynamic
relations which exhibit explicitly the reparametrization covariance (see in
appendix \ref{others}).

\section{Thermodynamical implications\label{some}}

The present proposal comes from by arising the reparametrization invariance to
a fundamental status within the microcanonical description. In the sake of
simplicity, let us analyze the implications of this symmetry in the
thermodynamical description of a Hamiltonian system with a microscopic
dynamics driven by short-range forces whose microcanonical description is
determined only from the consideration of \ the total energy $E$, that is, an
ordinary extensive system.

The preliminary interest of the present analysis is how identify the phase
transitions within the microcanonical description. Taking into account the
dynamical origin of this ensemble, a phase transition within the
microcanonical ensemble should be the macroscopic manifestation of certain
sudden change in the microscopic level which manifests itself as \textit{a
mathematical anomaly} of the Boltzmann entropy. Since the entropy is a
continuous scalar function, the most important mathematical anomalies of the
entropy per particle $s=S_{B}/N$ are the following:

\begin{itemize}
\item[(\textbf{A})] Regions where $s$ is not locally concave;

\item[(\textbf{B})] Every lost of analyticity in the thermodynamic limit
$N\rightarrow\infty$.
\end{itemize}

Anomaly type \textbf{A} is directly related with the first-order phase
transitions in the conventional Thermodynamics, while we shall show that the
anomaly type \textbf{B} can be associated to the second-order phase
transitions and other anomalous behaviors. We will concentrate in the next
subsections to the analysis of the anomalies type \textbf{A}.

\subsection{Why is so important the concavity of the microcanonical
entropy?\label{why}}

Most of applications of the Equilibrium Statistical Mechanics start from the
consideration of the Gibbs canonical ensemble:%
\begin{equation}
\hat{\omega}_{c}\left(  \beta,N\right)  =\frac{1}{Z\left(  \beta,N\right)
}\exp\left\{  -\beta\hat{H}_{N}\right\}  ,\label{gibbs}%
\end{equation}
which provides the macroscopic characterization of a Hamiltonian system with a
thermal contact with a heat bath. This description becomes equivalent to the
one carried out by using the microcanonical ensemble almost everywhere when
$N$ is large enough, which can be easily verified by considering the Laplace
transformation between their partition functions:
\begin{equation}
Z\left(  \beta,N\right)  =\int\exp\left(  -\beta E\right)  \Omega\left(
E,N\right)  dE.\label{LAP1}%
\end{equation}
This equation is conveniently rewritten by introducing the coarsed grained
entropy $S\left(  E,N\right)  =\ln\left\{  \Omega\left(  E,N\right)
\delta\varepsilon_{0}\right\}  $ and the Planck thermodynamic potential
$P\left(  \beta,N\right)  =-\ln Z\left(  \beta,N\right)  $\footnote{$P\left(
\beta,N\right)  $ is related with the Helmholtz free energy potential by
$F\left(  \beta,N\right)  =\beta^{-1}P\left(  \beta,N\right)  $.}:%

\begin{equation}
e^{-P\left(  \beta,N\right)  }=\delta\varepsilon_{0}^{-1}\int e^{-\left\{
\beta E-S\left(  E,N\right)  \right\}  }dE.
\end{equation}
The energy per particle dependence of the distribution function $\eta\left(
\varepsilon;\beta,N\right)  $ of the interest system within the canonical
ensemble is just the exponential function of the above integral: $\eta\left(
\varepsilon;\beta,N\right)  \propto\exp\left\langle -N\left(  \beta
\varepsilon-s\left(  \varepsilon,N\right)  \right)  \right\rangle $, being
$s\left(  \varepsilon;N\right)  =S\left(  N\varepsilon,N\right)  /N$ the
entropy per particle. When $N$ is large enough, $\eta\left(  \varepsilon
;\beta\right)  $ exhibits\ very sharp peaks around all those stationary points
$\left\{  \varepsilon_{m}\right\}  $ minimizing the functional $p\left(
\varepsilon;\beta,N\right)  =\beta\varepsilon-s\left(  \varepsilon;N\right)
$:
\begin{equation}
\beta=\frac{\partial s\left(  \varepsilon_{m};N\right)  }{\partial\varepsilon
}\text{ and }\kappa=\frac{\partial^{2}s\left(  \varepsilon_{m};N\right)
}{\partial\varepsilon^{2}}<0.\label{station}%
\end{equation}
However, in spite of the existence of several peaks, the presence of the
system size $N$ in the argument of the exponential dependence of the
distribution function $\eta\left(  \varepsilon;\beta,N\right)  $ allows the
Planck potential per particle to take asymptotically the value of the global
minimum $\varepsilon_{gm}$ of $p\left(  \varepsilon;\beta,N\right)  $:%
\begin{equation}
p\left(  \beta,N\right)  =\frac{P\left(  \beta,N\right)  }{N}\simeq
\beta\varepsilon_{gm}-s\left(  \varepsilon_{gm};N\right)  +O\left(  \frac{\ln
N}{N}\right)  ,\label{LT}%
\end{equation}
which determines in practice the thermodynamic values of the microscopic
observables within the canonical description.

There exist \textit{ensemble equivalence} when there is only one peak at a
given $\beta$ in the distribution function $\eta\left(  \varepsilon
;\beta,N\right)  $ (the macroscopic state at a given $\beta$ in the canonical
ensemble is equivalent to the macroscopic state of the system within the
microcanonical ensemble with $\varepsilon=\varepsilon_{gm}$). The square
dispersion of the energy $\left\langle \Delta\varepsilon^{2}\right\rangle
_{c}\simeq-1/\left(  N\kappa\right)  $ here decreases with the increasing of
the system size $N$, so that, the system energy is effectively fixed in the
thermodynamic limit. When $N$ is large but finite, the equivalence is only
asymptotic, that is, the canonical average $\left\langle \hat{A}\right\rangle
_{c}$ of any microscopic observable $\hat{A}$ converge asymptotically towards
the corresponding microcanonical average $\left\langle \hat{A}\right\rangle
_{m}$ with the $N$ increasing:
\begin{equation}
\delta_{A}\equiv\left\vert \left\langle \hat{A}\right\rangle _{c}-\left\langle
\hat{A}\right\rangle _{m}\right\vert \propto\frac{1}{N}.
\end{equation}
The reader may notice that the equation \ref{LT} is just the \textit{Legendre
transformation} $P\left(  \beta,N\right)  =\beta E-S\left(  E,N\right)  $
which relates the thermodynamic potentials of the corresponding ensembles,
whose validity justifies the applicability of the well-known thermodynamic
formalism of the conventional Thermodynamics.

On the other hand, there exist \textit{ensemble inequivalence} when the
distribution function $\eta\left(  \varepsilon;\beta,N\right)  $ exhibits a
multimodal character, being this feature a direct manifestation of the
existence of several metastable "microcanonical states" within the canonical
description at a given $\beta$. The energy interchange with the heat bath
(thermostat) provokes here a random transition of the system energy among
these metastable states, and consequently, the system undergoes the incidence
of \ very large fluctuations and non homogeneities.

Undoubtedly, this is the well-known phenomenon of \textit{phase coexistence}
usually referred as the occurrence of a \textit{first-order phase transition}:
any small variation of the thermostat parameter $\beta$ close to a critical
value $\beta_{C}$ provokes that certain metastable state becomes in a global
minima of the functional $p\left(  \varepsilon;\beta,N\right)  $, leading in
this way to the occurrence of an abrupt change of the average energy per
particle $\left\langle \varepsilon\right\rangle _{c}$\ of the system. Since
the canonical average of the energy per particle can be derived from the
Planck potential per particle $p\left(  \beta;N\right)  $ as follows:
\begin{equation}
\left\langle \varepsilon\right\rangle _{c}=-\frac{1}{NZ\left(  \beta,N\right)
}\frac{\partial Z\left(  \beta,N\right)  }{\partial\beta}\equiv\frac{\partial
p\left(  \beta;N\right)  }{\partial\beta},
\end{equation}
such thermodynamic function exhibits a sudden change in its first derivative
at the critical point $\beta_{C}$, which becomes a discontinuity in the
thermodynamic limit, being this behavior a feature of the first-order phase
transitions in the conventional Thermodynamics. \ 

Since the transition from a given metastable state to another demands the
incidence of very large fluctuations of the total energy, such events involve
a very large characteristic time due to the probability of occurrence of a
very large fluctuation decreases exponentially with the increasing of the
system size $N$. This dynamical behavior leads to an exponential \ divergence
of the correlations times with $N$ during the Monte Carlo simulations based on
the Gibbs canonical ensemble, a phenomenon referred as a \textit{supercritical
slowing down} \cite{wang2}.

Supposing the existence and continuity of the first and the second derivatives
of the entropy, the existence of two or more stationary points satisfying the
conditions \ref{station}\ at a given $\beta$ is also associated to the
presence of convex regions of the microcanonical entropy.\ Such convex regions
represent thermodynamic states with a \textit{negative heat capacity }within
the microcanonical description:%
\begin{equation}
C^{\left(  m\right)  }=\left(  \frac{dT}{dE}\right)  ^{-1}\equiv-\left(
\frac{\partial S}{\partial E}\right)  ^{2}\left(  \frac{\partial^{2}%
S}{\partial E^{2}}\right)  ^{-1},
\end{equation}
which is an anomalous behavior since the heat capacity in the canonical
ensemble is always positive:%
\begin{equation}
C^{\left(  c\right)  }=\frac{dE}{dT}=\beta^{2}\left\langle \left(  \Delta
E\right)  ^{2}\right\rangle _{c}\geq0.
\end{equation}

According to the conditions \ref{station}, such anomalous thermodynamic states
will not be observed in the Gibbs canonical ensemble \ref{gibbs}.
Particularly, the corresponding microcanonical entropy can not be obtained
from the Legendre transformation $S=\beta E-P$ starting from the Planck
potential $P$. Gross refers this situation\ as a \textit{cathastrophe} of the
Legendre transformation \cite{gro1}. Thus, the existence of an ensemble
inequivalence associated to the convexity of the microcanonical entropy leads
to a significant lost of information when the canonical description is
performed instead the microcanonical description: the Gibbs canonical ensemble
is unable to describe the system features during the phase coexistence, i.e.
the existence of a non vanishing interphase tension \cite{gro na}.

\textit{Summarizing}: The concavity of the Boltzmann entropy is a necessary
condition for the ensemble equivalence and the validity of the Legendre
transformation $P=\beta E-S$ between the thermodynamic potentials which
supports the applicability of the well-known thermodynamic formalism in the
conventional Thermodynamics. The existence of convex regions in the Boltzmann
entropy of short-range interacting systems can be related with the existence
of the \textit{first-order phase transitions}, which are also associated with
the lost of analyticity of the Planck $P\left(  \beta,N\right)  $ or the
Helmholtz free energy $F\left(  \beta,N\right)  $\ potential (discontinuity of
first derivatives) in the thermodynamic limit. The existence of the multimodal
character of the canonical energy distribution function under the presence of
first-order phase transitions is the origin of the phenomenon of critical
slowing down in the neighborhood of the critical point during the Monte Carlo
simulations based on the Gibbs canonical ensemble \ref{gibbs}.

\subsection{The concave or convex character of the microcanonical entropy is
an ambiguous concept from the reparametrization invariance
viewpoint!\label{amb}}

As already discussed in the subsection above, the concavity of the
microcanonical entropy is a very important condition within the conventional
Thermodynamics. However, the consideration of the reparametrization invariance
leads naturally to the \textit{ambiguous character of the concavity of the
microcanonical entropy}: the concave or convex character of a scalar function
like the entropy depends crucially on the parametrization used for describe
it. Let us consider a trivial example.

Let $s$ be a positive real map defined on a seminfinite Euclidean line
$\mathcal{L}$, $s:\mathcal{L}\rightarrow R^{+}$, which is given by the concave
function $s\left(  x\right)  =\sqrt{x}$ in the coordinate representation
$\mathcal{R}_{x}$ of\ $\mathcal{L}$ (where $x>0$). Let $\varphi$ be a
reparametrization change $\varphi:\mathcal{R}_{x}\rightarrow\mathcal{R}_{y}$
given by $y=\varphi\left(  x\right)  =x^{\frac{1}{4}}$. The map $s$ in the new
representation $\mathcal{R}_{y}$ of the seminfinite Euclidean line
$\mathcal{L}$ is now given by the function $s\left(  y\right)  =y^{2}$ (with
$y>0$), which is clearly a convex function.

The ambiguous character of the microcanonical entropy is straightforwardly
followed from the non reparametrization invariance of the entropy Hessian
shown in the equation \ref{hessian_tr}. The only exceptions are those
thermodynamic states where the corresponding entropy shows a local extreme:
the first derivatives of the entropy vanish there and the Hessian behaves
\textit{eventually} as a second rank tensor. An analogue situation was used by
Ruppeiner\ for establish a Riemannian interpretation of the Thermodynamics in
the ref.\cite{rupper} (see in appendix \ref{ruppeiner}).

Reader may object correctly that such ambiguity of the concave character of
the entropy is irrelevant within the conventional Thermodynamics because of
the only one \textit{admissible} representation $\Theta\left(  E\right)  $ in
this framework is the total energy, $\Theta\left(  E\right)  \equiv E$. This
viewpoint follows directly from the fact that most of applications of the
Equilibrium Statistical Mechanics start from the consideration of the Gibbs
canonical ensemble \ref{gibbs} (or more general, from the
\textit{Boltzmann-Gibbs distributions}). However, it is necessary to recall
that the reparametrization invariance is only relevant in the microcanonical
description since the Gibbs canonical ensemble \ref{gibbs} is not
reparametrization invariant. Their different background physical conditions
are crucial: the microcanonical ensemble describes an isolated system while
the Gibbs canonical ensemble describes the same system with a weak interaction
with a thermostat, that is, an open system.

At first glance, the above ambiguity suggests us that \textit{the first-order
phase transitions associated with the existence of convex regions of the
entropy could not be a phenomenon microcanonically} \textit{relevant due to
the reparametrization invariance of this ensemble}. This preliminary idea
contradicts our common sense educated in the everyday practice where the phase
coexistence is a very familiar phenomenon. Nevertheless, most of these
experiences are observed under those background conditions which lead to the
applicability of the Gibbs canonical ensemble \ref{gibbs}. We shall
demonstrate in the next subsection that this ensemble can be generalized in
order to develop a thermodynamic framework where the reparametrization
invariance plays a more fundamental role than in the conventional Thermodynamics.

\subsection{Generalized canonical ensemble\label{gen.can.ens}}

The Gibbs canonical ensemble \ref{gibbs} belongs to a family of equilibrium
distribution functions called the \textit{Boltzmann-Gibbs distributions}
\cite{gallavotti}. Generally speaking, the Boltzmann-Gibbs distributions
provide the macroscopic characterization of an open system in thermodynamic
equilibrium with its very large surrounding, where this system and its
surrounding constitute a \textit{closed environment}. Of course, the specific
form of this distribution functions depends on the external conditions which
are been imposed to the interest system, i.e.: the well-known Gran canonical
ensemble:%
\begin{equation}
\hat{\omega}\left(  \beta,\mu\right)  =\frac{1}{Z\left(  \beta,\mu\right)
}\exp\left\{  -\beta\left(  \hat{H}+\mu\hat{N}\right)  \right\}  ,
\end{equation}
describes a system which is able to interchange energy and particles with its
surrounding (reservoir).

Although the Boltzmann-Gibbs distribution functions play a fundamental role in
most of applications of the Statistical Mechanics and the classical
fluctuation theory, they are not the only physically admissible: an open
system could be arbitrarily affected by many other external conditions so that
the whole experimental setup (the interest system + its surrounding) do not
correspond necessarily to a closed environment. It is important to mention
that although an open system should exhibit a stationary macroscopic state
when the external conditions are also stationary, most of these stationary
states do not correspond to \textit{equilibrium conditions }since the
probability currents between microstates may not vanish.

Let us consider certain experimental setup which allows the interest
short-range Hamiltonian system $\hat{H}_{N}$ to interchange energy with its
surroundings in some stationary way that leads this system to a thermodynamic
equilibrium described by the following \textit{generalized canonical
ensemble}:%
\begin{equation}
\hat{\omega}_{GC}\left(  \eta,N\right)  =\frac{1}{Z\left(  \eta,N\right)
}\exp\left\langle -\eta\hat{\Theta}_{N}\right\rangle , \label{g.canonical}%
\end{equation}
where $\hat{\Theta}_{N}=\Theta\left(  \hat{H}_{N}\right)  $ corresponds to
certain reparametrization of the total energy $E\rightarrow\Theta
=\Theta\left(  E\right)  =N\varphi\left(  E/N\right)  $ preserving the
extensive character of the energy in the thermodynamic limit. Obviously, the
generic form of the distribution function \ref{g.canonical} contains the Gibbs
canonical ensemble \ref{gibbs} as a particular case, where the only explicit
difference is the introduction of the energy reparametrization $\Theta\left(
E\right)  $.

The generalized distribution function \ref{g.canonical} can be straightforward
derived from the maximization of the Shannon-Boltzmann-Gibbs extensive entropy
\ref{sbg} by preserving the average value of the microscopic observable
$\Theta\left(  E\right)  $:%
\begin{equation}
\left\langle \Theta\right\rangle =\sum_{k}\Theta\left(  E_{k}\right)  p_{k},
\label{average}%
\end{equation}
This observation allows us to understand that the role of the experimental
setup which leads to the generalized canonical ensemble \ref{g.canonical} is
just to preserve the average \ref{average} instead the energy average:%
\begin{equation}
\left\langle E\right\rangle =\sum_{k}E_{k}p_{k}.
\end{equation}
Obviously, the present experimental setup is a more sophisticate version of
the thermostat of the Gibbs canonical ensemble which keeps fixed the
"canonical" parameter $\eta$ instead the inverse temperature $\beta$. We shall
refer this external setup as a \textit{generalized thermostat} whose canonical
parameter $\eta$ exhibits in the present framework the same thermodynamic
relevance of the ordinary temperature in the conventional Thermodynamics.

The generalized canonical ensemble defines a complete family of equilibrium
distribution functions which exhibit the same features of the Gibbs canonical
ensemble almost everywhere. This affirmation follows directly from the
reparametrization invariance of the microcanonical ensemble:%
\begin{equation}
\Omega\left(  E,N\right)  dE=\Omega\left(  \Theta,N\right)  d\Theta,
\end{equation}
which allows us to express the partition function of the generalized canonical
ensemble (\ref{g.canonical}) by using the following Laplace Transformation:
\begin{equation}
Z\left(  \eta,N\right)  =\int\exp\left(  -\eta\Theta\right)  \Omega\left(
\Theta,N\right)  d\Theta. \label{LAP2}%
\end{equation}
The analogy between the equations \ref{LAP2} and \ref{LAP1} allows us to
understand that the generalized canonical ensemble admits a simple extension
of the features of the Gibbs canonical ensemble by using the reparametrization
$\Theta\left(  E\right)  $ instead the total energy $E$ of the extensive system.

Particularly, the generalized Planck potential $P\left(  \eta,N\right)  =-\ln
Z\left(  \eta,N\right)  $ in the thermodynamic limit $N\rightarrow\infty$ can
be approximated by the global minimum $\Theta_{gm}\in\left\{  \Theta
_{m}\right\}  $\ of the Legendre transformation:%
\begin{equation}
P\left(  \eta,N\right)  \simeq\eta\Theta_{gm}-S\left(  \Theta_{gm},N\right)  ,
\end{equation}
whose the stationary points $\left\{  \Theta_{m}\right\}  $ are derived from
the conditions:%
\begin{equation}
\eta=\frac{\partial S\left(  \Theta_{m},N\right)  }{\partial\Theta}\text{ and
}\frac{\partial^{2}S\left(  \Theta_{m},N\right)  }{\partial\Theta^{2}}<0.
\label{station2}%
\end{equation}
Thus, such generalized canonical description at a given $\eta$ becomes
equivalent in the thermodynamic limit $N\rightarrow\infty$ to a macroscopic
state of the microcanonical ensemble when the global minimum $\Theta_{gm}$ is
the only one stationary point. We can refer this situation as a \textit{local
ensemble equivalence} between a given generalized canonical ensemble and the
microcanonical ensemble. We can also refer to a \textit{global ensemble
equivalence} when these ensembles are locally equivalent everywhere.

Two macroscopic states corresponding to different generalized canonical
descriptions (with reparametrizations $\Theta_{1}\left(  E\right)  $ and
$\Theta_{2}\left(  E\right)  $ respectively) are \textit{mutually equivalent}
when these macroscopic states are locally equivalent to the same macroscopic
state in the microcanonical ensemble. It is very easy to derive from the
stationary conditions \ref{station2} that the canonical parameters
$\eta_{\Theta_{1}}$ and $\eta_{\Theta_{2}}$ of mutually equivalent macroscopic
states are related by the following transformation rule:%
\begin{equation}
\eta_{\Theta_{2}}=\frac{\partial S}{\partial\Theta_{2}},\eta_{\Theta_{1}%
}=\frac{\partial S}{\partial\Theta_{1}}\Rightarrow\eta_{\Theta_{2}}%
=\frac{\partial\Theta_{1}}{\partial\Theta_{2}}\eta_{\Theta_{1}}.
\label{transf_2}%
\end{equation}
This is exactly the transformation rule during the reparametrization changes
of unidimensional covariant vectors expressed in the equation
\ref{covariant_vector}. While a microcanonical state remains invariable under
the reparametrization changes, this kind of transformations provokes the
transition among macroscopic states associated to different generalized
canonical ensembles \ref{g.canonical} which could non necessarily be mutually
equivalent among them.

The transformation rule \ref{transf_2} allows us a better understanding about
the nature of the generalized thermostat associated to a generalized canonical
ensemble by compared this last one with an ordinary thermostat during their
interaction with a very large system. During a reparametrization change from a
generalized canonical ensemble with reparametrization $\Theta\left(  E\right)
$ towards the Gibbs canonical ensemble the respective canonical parameters
$\eta$ and $\beta$ are related as follows:
\begin{equation}
\beta\left(  E;\eta\right)  =\frac{\partial\Theta\left(  E\right)  }{\partial
E}\eta. \label{fluct.temp}%
\end{equation}
This equation tells us that a generalized thermostat with fixed canonical
parameter $\eta$ can be taken as an ordinary thermostat whose inverse
temperature $\beta$ depends on the instantaneous value of the total energy
$E$\ of the system under analysis. Therefore, \textit{the total energy }%
$E$\textit{ and the inverse temperature }$\beta$\textit{ experience correlated
fluctuations in the generalized canonical ensemble} \ref{g.canonical}. This
fact is a remarkable feature of the generalized canonical ensemble in regard
with the Gibbs canonical ensemble where $\beta$ is fixed and $E$ fluctuates
around its mean value $\left\langle E\right\rangle $. We shall shown in the
subsection \ref{unc.rel}\ that the characterization of the fluctuations of the
energy $E$\ and the inverse temperature $\beta$ demands a suitable extension
of the fluctuation theory of the conventional Thermodynamics to the present framework.

\subsection{Generalized Metropolis Monte Carlo method\label{gmmc}}

The basic problem in equilibrium statistical mechanics is to compute the phase
space average, in which Monte Carlo method plays a very important role
\cite{mc1}. Among all admissible statistical ensembles used with the above
purpose, the microcanonical ensemble provides the most complete
characterization of a given system in thermodynamic equilibrium. However,
microcanonical calculations could be difficult to carry out directly in a
given application. A very simple way to overcome this difficulty is to
consider the relationship among the microcanonical description of a large
enough system with other statistical ensembles (like the Gibbs canonical
ensemble \ref{gibbs} or the multicanonical ensemble \cite{berg1,berg2}).

Suitable estimates of the microcanonical averages could be easily obtained
from the Metropolis importance sample algorithm (MMC) \cite{met}\ by using the
equivalence between the microcanonical ensemble and the Gibbs canonical
ensemble. This Monte Carlo method has some important features. It is extremely
general and each moves involves $O\left(  1\right)  $ operations. However, its
dynamics suffers from \textit{critical slowing down,} that is, if $N$ is the
system size, the correlation time $\tau$ diverges as a critical temperature is
approached: the divergence follows a power law behavior $\tau\propto N^{z}$ in
second-order phase transitions, while $\tau$ diverges exponentially with $N$
in first-order phase transitions as consequence of the ensemble inequivalence
\cite{wang2}.

The critical slowing down observed in the neighborhood of the critical point
of the second-order phase transitions is intimately related with the existence
of the long-range order (divergence of correlation length) which characterizes
the microscopic picture of this kind of phase transitions. This difficulty can
be overcome in many physical systems by using appropriated clusters algorithms
\cite{wolf}.

As already explained in the subsection \ref{why}, the supercritical slowing
down (exponential divergence of the correlation times) of the first-order
phase transitions is originated from the ensemble inequivalence of the
canonical description. Such difficulty is usually overcome by using the
multicanonical ensemble \cite{berg1,berg2}, a methodology which reduces the
exponential divergence of the correlation times with respect to system size to
a power in the first-order phase transitions. The multicanonical ensemble
flattens out the energy distribution, which allows the computation of the
density of states $\Omega\left(  E,N\right)  $ for all values of $E$ in only
one run.

An alternative way to overcome the supercritical slowing down of the MMC
algorithm in the neighborhood of the first-order phase transitions is to ovoid
the underlying ensemble inequivalence by using the generalized canonical
ensemble \ref{g.canonical}. The method, the generalized canonical Metropolis
Monte Carlo (GCMMC) is explained in details in the ref.\cite{vel-mmc}, so
that, we shall limit in this subsection to expose its more important features.

Loosely speaking, the GCMMC algorithm is essentially the same MMC algorithm
with the generalized canonical weight $\omega_{GC}\left(  E\right)
=\exp\left[  -\eta\Theta\left(  E\right)  \right]  $ instead of the usual
Gibbs\ canonical weight $\omega_{c}\left(  E\right)  =\exp\left[  -\beta
E\right]  $. The probability $p$ for the acceptance of a Metropolis move is
given now by $p=\min\left\{  1,\exp\left[  -\eta\Delta\Theta\right]  \right\}
$ where $\Delta\Theta=\Theta\left(  E+\Delta\varepsilon\right)  -\Theta\left(
E\right)  $. Since $\Delta\varepsilon<<E$ when the interest system is large
enough, we can use the approximation $\eta\Delta\Theta\simeq\left\{
\eta\partial\Theta\left(  E\right)  /\partial E\right\}  \Delta\varepsilon
\equiv\beta\left(  E;\eta\right)  \Delta\varepsilon$, which allows us to
rewrite the acceptance probability $p$ as follows:%
\begin{equation}
p\simeq\min\left\{  1,\exp\left[  -\beta\left(  E;\eta\right)  \Delta
\varepsilon\right]  \right\}  .
\end{equation}
The reader can recognize the natural appearance of the fluctuating inverse
temperature $\beta\left(  E;\eta\right)  $ characterizing the generalized
thermostat with parameter $\eta$ and energy reparametrization $\Theta\left(
E\right)  $ introduced in the equation \ref{fluct.temp}. This feature of the
GCMMC algorithm allows the Metropolis dynamics to explore regions which are
inaccessible for any other Monte Carlo method based on the Gibbs canonical
ensemble, that is, regions which are characterized by exhibiting a negative
heat capacity (see in subsection \ref{why}).

It is very easy to show why and when the GCMMC algorithm works. Considering
the following expression of the generalized partition function:%

\begin{equation}
Z\left(  \eta,N\right)  =\frac{1}{\delta\varepsilon_{0}}\int\exp\left[
-\eta\Theta\left(  E\right)  +S\left(  E,N\right)  \right]  dE,
\end{equation}
the application of the steepest descent method up to the Gaussian
approximation allows us to express the conditions of the local ensemble
equivalence between the macrostate of the generalized canonical ensemble
\ref{g.canonical} with canonical parameter $\eta$ and a microcanonical
macrostate with total energy $E_{e}$ as follows:
\begin{equation}
\beta\left(  E_{e}\right)  =\frac{\partial S\left(  E_{e},N\right)  }{\partial
E}=\beta\left(  E_{e};\eta\right)  =\eta\frac{\partial\Theta\left(  E\right)
}{\partial E}, \label{c1}%
\end{equation}%
\begin{equation}
\left\langle \delta E^{2}\right\rangle ^{-1}=\frac{1}{N\sigma_{\varepsilon
}^{2}}\equiv\eta\Theta^{\prime\prime}\left(  E_{e}\right)  -S^{\prime\prime
}\left(  E_{e},N\right)  >0. \label{c2}%
\end{equation}

The condition \ref{c1} identifies what it is usually referred as the
microcanonical inverse temperature $\beta\left(  E\right)  =\partial S\left(
E,N\right)  /\partial E$ \ with the value of the effective inverse temperature
$\beta\left(  E;\eta\right)  $ of the generalized thermostat at the stationary
point $E_{e}$. Reader may notice that this expression is just the
transformation rule \ref{transf_2}. The positive definition of the energy
square dispersion $\left\langle \delta E^{2}\right\rangle $ expressed in the
equation \ref{c2} imposes a restriction to the energy reparametrization
$\Theta\left(  E\right)  $. It is very easy to show by combining the equations
\ref{c1} and \ref{c2} that the positive definition of the energy square
dispersion $\left\langle \delta E^{2}\right\rangle $ is ensured by the
concavity of the microcanonical entropy in the reparametrization $\Theta$:%
\begin{equation}
\left\langle \delta E^{2}\right\rangle =-\left(  \frac{\partial\Theta\left(
E\right)  }{\partial E}\right)  ^{2}\frac{\partial^{2}S}{\partial\Theta^{2}%
}\Rightarrow\frac{\partial^{2}S}{\partial\Theta^{2}}<0.
\end{equation}

We can used in principle several generalized canonical ensembles in order to
describe all admissible macrostates within the microcanonical description.
According to the ref.\cite{vel-mmc}, we can use the Gibbs canonical ensemble
to describe all those energetic regions where the local ensemble equivalence
takes place, while consider a generalized canonical ensemble with
reparametrization $\Theta\left(  E\right)  =N\varphi\left(  E/N\right)  $:%
\begin{equation}
\frac{\partial\varphi\left(  \varepsilon\right)  }{\partial\varepsilon}%
=\exp\left\{  -\lambda\left(  \varepsilon_{2}-\varepsilon\right)  \right\}  ,
\end{equation}
within an anomalous region $\left(  \varepsilon_{1},\varepsilon_{2}\right)  $,
with $\lambda^{-1}\simeq\beta_{c}$, being $\varepsilon=E/N$ the energy per
particle, and $\beta_{c}$, an estimation of the inverse critical temperature
of the first-order phase transitions within the Gibbs canonical ensemble.

\subsection{Thermodynamic uncertainly relation\label{unc.rel}}

As already discussed, the total energy of the interest system $E$\ and the
inverse temperature of the generalized thermostat $\beta\left(  E,\eta\right)
$ exhibit correlated fluctuations within the generalized canonical ensemble.
Since the relative energy fluctuations $\overline{\delta E}/E\sim1/N$ (with
$\overline{\delta E}=\sqrt{\left\langle \delta E^{2}\right\rangle }$) drops to
zero in the thermodynamic limit $N\rightarrow\infty$, their corresponding
microcanonical values in this limit are just the average expectation values
within the generalized canonical ensemble $E_{e}=\left\langle E\right\rangle $
and $\beta\left(  E_{e}\right)  =\left\langle \beta\left(  E,\eta\right)
\right\rangle $.

The correlation function $\left\langle \delta\beta\delta E\right\rangle $
between the fluctuations of the effective inverse temperature $\beta\left(
E;\eta\right)  $ of the generalized thermostat and the fluctuations of the
total energy of the interest system $\delta E$ can be obtained as follows:
\begin{equation}
\delta\beta\simeq\eta\Theta^{\prime\prime}\left(  E_{e}\right)  \delta
E\Rightarrow\left\langle \delta\beta\delta E\right\rangle =\eta\Theta
^{\prime\prime}\left(  E_{e}\right)  \left\langle \delta E^{2}\right\rangle ,
\end{equation}
which can be rephrased conveniently by eliminating the factor $\eta
\Theta^{\prime\prime}\left(  E\right)  $ using the equation \ref{c2}. The
resulting expression:
\begin{equation}
\left\langle \delta\beta\delta E\right\rangle =1+\left\langle \delta
E^{2}\right\rangle \frac{\partial^{2}S\left(  E_{e},N\right)  }{\partial
E^{2}}, \label{thur1}%
\end{equation}
is a very remarkable equation which could be referred as a
\textit{thermodynamic uncertainly relation}.\ 

The reader can notice that the result \ref{thur1} does not makes explicit
reference to the reparametrization $\Theta\left(  E\right)  $, so that, it
presents \textit{a general applicability for a large system in thermodynamic
equilibrium interchanging energy with its surrounding}. This identity
represents a restriction among the fluctuations of the energy of the interest
system $E$ and the inverse temperature $\beta$ of a given generalized
thermostat, providing in this way a suitable generalization of the
\textit{classical fluctuation theory }\cite{rupper}. Since the total energy
$E$ is just an admissible reparametrization for the microcanonical
description, the reparametrization invariance supports without any lost of
generality the following extension of the identity \ref{thur1}:
\begin{equation}
\left\langle \delta\eta\delta\Theta\right\rangle =1+\left\langle \delta
\Theta^{2}\right\rangle \frac{\partial^{2}S}{\partial\Theta^{2}}.
\label{thur2}%
\end{equation}

While effective inverse temperature of the generalized thermostat can be fixed
($\delta\beta=0$) whenever the entropy be a concave function in terms of the
total energy $E$, the inverse temperature dispersion $\delta\beta$ can not
vanish when the entropy be a convex function. All those anomalous behaviors
appearing in the Gibbs canonical ensemble during the ensemble inequivalence
are also related with the downfall of the well-known Gaussian estimation of
the energy dispersion $\left\langle \delta E^{2}\right\rangle =-\left\{
\partial^{2}S\left(  E,N\right)  /\partial E^{2}\right\}  ^{-1}$ in such
anomalous regions. There $\partial^{2}S\left(  E,N\right)  /\partial E^{2}%
\geq0$, and consequently, $\left\langle \delta\beta\delta E\right\rangle
\geq1$. A better analysis allows us to obtain the inequalities $\left\langle
\delta E^{2}\right\rangle \leq-\left\{  \partial^{2}S\left(  E,N\right)
/\partial E^{2}\right\}  ^{-1}$ whenever entropy be a concave function, while
$\left\langle \delta\beta^{2}\right\rangle \geq\partial^{2}S\left(
E,N\right)  /\partial E^{2}$ when entropy is convex.

According to the identity \ref{thur1}, the total energy $E$ and the inverse
temperature $\beta$ are \textit{complementary thermodynamic quantities}, in
complete analogy like the time $\tau$ and the energy $E$\ or the coordinates
$q$\ and the corresponding momentum $p$ are complementary quantities in the
Quantum Mechanics ($\delta\tau\delta E\geq\hbar$, $\delta q\delta p\geq\hbar
$). Although there is nothing strange in the existence of this kind of
relationship between temperature and the energy\footnote{Within the Quantum
Mechanics framework the Gibbs canonical weight $\hat{\omega}\left(
\beta\right)  =\exp\left(  -\beta\hat{H}\right)  $ can be considered as a
evolution operator $\hat{S}\left(  \tau\right)  =\exp\left(  -i\tau\hat
{H}/\hbar\right)  $ with \textit{imaginary time} $\tau\rightarrow-i\beta\hbar
$.}, the conventional Thermodynamics usually deals with equilibrium situations
where temperature is fixed and the energy fluctuates (Boltzmann-Gibbs
ensemble), or the energy be fixed and the temperature fluctuates
(microcanonical ensemble). The generalized canonical ensemble
\ref{g.canonical} provides an appropriate framework for dealing with
equilibrium situations where both energy and temperature are not fixed.

The thermodynamic uncertainly relation \ref{thur1} admits the following
generalization when the open system is controlled by several thermodynamical
parameters:
\begin{equation}
\left\langle \delta\beta_{k}\delta I^{m}\right\rangle =\delta_{k}%
^{m}+\left\langle \delta I^{n}\delta I^{m}\right\rangle \frac{\partial^{2}%
S}{\partial I^{n}\partial I^{k}},
\end{equation}
expression with provides a new relevance of the entropy Hessian within the
present generalization of the classical fluctuation theory. The development of
the present ideas deserves a further study.

\subsection{The role of the external conditions\label{irre}}

The reader can noticed by a simple inspection of the stationary conditions
\ref{station2}\ that the local ensemble equivalence demands the local
concavity of the microcanonical entropy, while the global ensemble equivalence
demands the global concave character of this thermodynamic potential. As
already illustrated in the subsections \ref{amb} and \ref{gen.can.ens}, such
conditions depend crucially on the nature of the reparametrization
$\Theta\left(  E\right)  $ used in the generalized canonical ensemble
\ref{g.canonical}.\ 

Obviously, regions of ensemble inequivalence also depends on the
reparametrization. Since different reparametrizations in the generalized
canonical ensemble \ref{g.canonical} represent different experimental setups
(the using of different generalized thermostats), the existence of all those
anomalous behaviors associated to the ensemble inequivalence depend crucially
on the nature of the experimental setup. Therefore, we can not refer in the
present framework to the existence of an ensemble equivalence without
indicating the external conditions which have been imposed to the interest
system. This situation is completely analogue to the question about the
character of the motion of a given particle without specifying the reference frame.

As already pointed in the subsection \ref{why}, the phenomenon of phase
coexistence associated to the ensemble inequivalence is interpreted as a
first-order phase transition in the conventional Thermodynamics. However, the
ensemble inequivalence represents an anomaly within the canonical description
which reflects the inability of this ensemble in considering all those
macrostates observed in the microcanonical description of the interest system.
Consequently, what we called \textit{a first-order phase transition in the
conventional Thermodynamics is a phenomenon only relevant in the Gibbs
canonical description: such behavior does not represent anything anomalous
within the microcanonical description}.

Obviously, the presence of an ensemble inequivalence in the Gibbs canonical
description can be easily recognized within the microcanonical ensemble by the
existence of convex regions of the entropy when this description is performed
in terms of the total energy $E$. However, it is necessary to emphasize that
there is nothing anomalous in this behavior of the entropy since the
microcanonical description is reparametrization invariance, and consequently,
\textit{any anomaly within this ensemble must be also reparametrization
invariant}. Therefore, the convex or concave character of the entropy can not
be used as an indicator of real anomaly within the microcanonical description
since this mathematical property can be arbitrarily modified with a simple
change of the reparametrization $E\rightarrow\Theta\left(  E\right)  $.

It is convenient to recall one time again the generic
(experimental)\ definition of phase transition: " ... a phase transition
brings about a sudden change of the macroscopic properties of a system while
smoothly varying of a \textit{control parameter}". To be more precise, we need
to add that the behavior of the system is affected by the way its
thermodynamic equilibrium is controlled by the experimental setup, and
consequently, the thermodynamic anomalies which could be appearing under
certain external conditions may not observed by using other experimental
setup. Furthermore, taking into account the example of the Gibbs canonical
ensemble which is unable to describe many thermodynamic states during the
occurrence of a first order phase transition, it is very important to remark
that \textit{other relevant thermodynamic anomalies could be hidden behind of
any lost of information associated to the ensemble inequivalence}.

\section{A paradigmatic example\label{potts}}

\subsection{Hamiltonian of the Potts model}

The present section will be devoted to perform the thermostatistical
description of a simple model exhibiting phase transitions in order to
illustrate some aspects already discussed in the above section. Let us began
the present section by introducing the Potts model \cite{pottsm}, a toy model
whose Hamiltonian is defined as:%

\begin{equation}
H=\sum_{\left(  i,j\right)  }\left\{  1-\delta_{\sigma_{i},\sigma_{j}%
}\right\}  , \label{definition}%
\end{equation}
on a hypercubic dimensional lattice, being $\sigma_{i}$ the spin state at the
\textit{i-th} lattice point which can take $q\left(  \geq2\right)  $ possible
integer values or components, $\sigma_{i}=1,~2,...q$, where the sum is over
pairs of nearest neighbor lattice points. Hereafter we will assume periodic
boundary conditions. It is very easy to notice that the Potts model with $q=2$
is equivalent to the Ising model of the ferromagnetism. Obviously, the cases
with $q>2$ are suitable generalizations of this paradigmatic toy model which
also admit a ferromagnetic interpretation (see in subsection \ref{magnet}).

Potts models are very amenable for studying phase transitions with different
order, i.e. $q=2$ (Ising model) exhibits a continuous (second-order) phase
transition, while $q=10$ shows a discontinuous (first-order) phase transition.
The present section will be devoted to study the main thermodynamic features
of the $q=10$ Potts model on a $L\times L$ square lattice by using the GCMMC
algorithm described in the subsection \ref{gmmc} and the ref.\cite{vel-mmc}.

\subsection{Thermodynamical potentials for $q=10$}

The $q=10$ states Potts model exhibits a first-order phase transition
associated to the ensemble inequivalence, which provokes the existence of a
supercritical slowing down during the ordinary Metropolis dynamics based on
the consideration of the Gibbs canonical ensemble. Clusters algorithms, like
Swendsen-Wang or Wolf algorithms \cite{wolf}, do not help in this case due to
they are still based on the consideration of the canonical ensemble
\cite{gore}, and consequently, the supercritical slowing down associated to
the ensemble inequivalence persists \cite{wang2}.

\begin{figure}
[t]
\begin{center}
\includegraphics[
height=3.5129in,
width=3.2119in
]%
{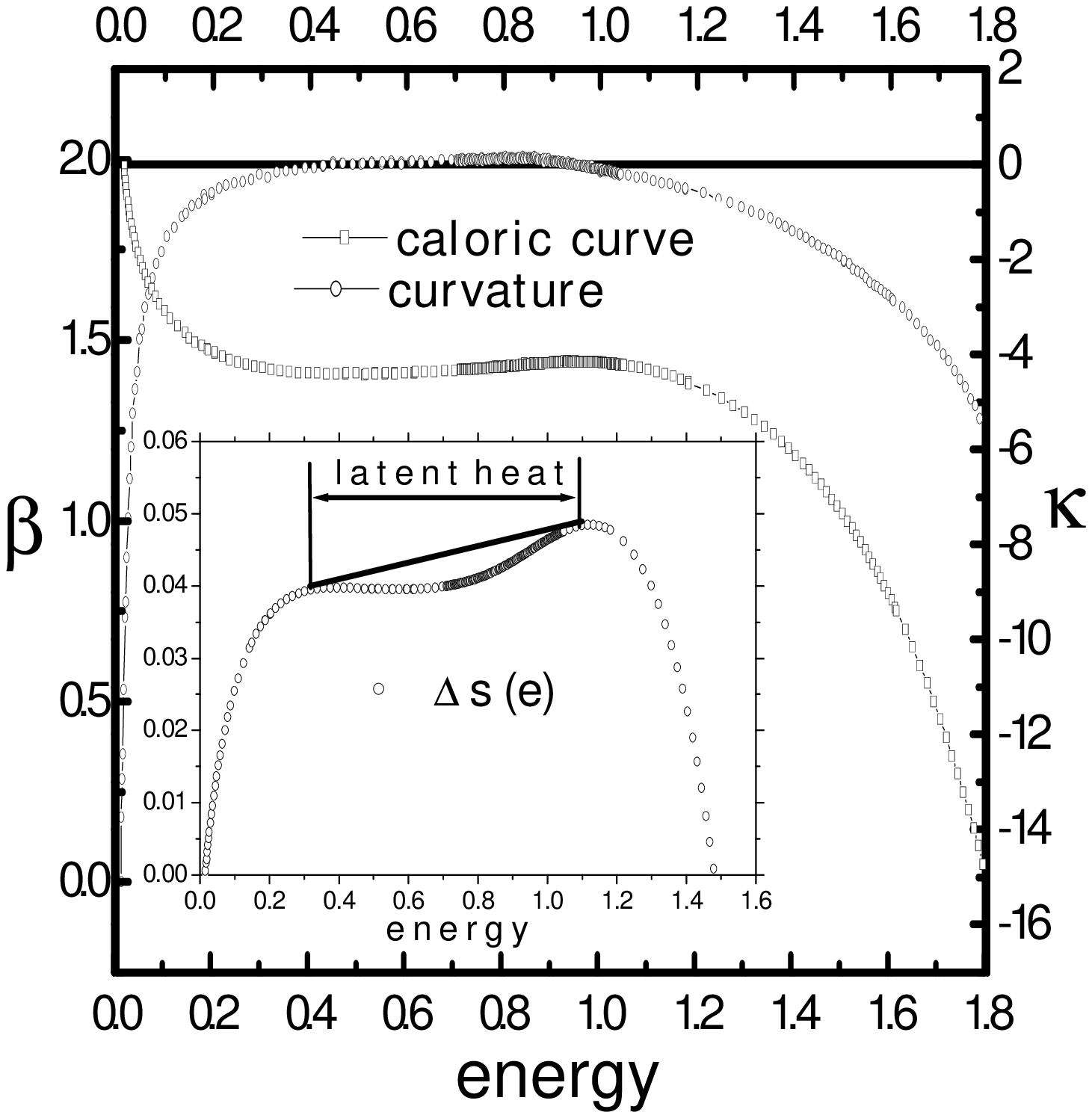}%
\caption{Microcanonical caloric curve $\beta\left(  \varepsilon\right)
=\partial s\left(  \varepsilon\right)  /\partial\varepsilon$ and curvature
$\kappa\left(  \varepsilon\right)  =\partial^{2}s\left(  \varepsilon\right)
/\partial\varepsilon^{2}$ of the $25\times25$ Potts model with $q=10$ states
with periodic boundary conditions obtained by using the GCMMC algorithm.
Notice the energetic region with a negative heat capacity. The inserted graph
shows the behavior of the relative microcanonical entropy $\Delta s\left(
\varepsilon\right)  $ in which is remarkable the existence of a \textit{convex
intruder}. Errors are smaller than the symbols linear dimension.}%
\label{thermo1.eps}%
\end{center}
\end{figure}

\begin{figure}
[t]
\begin{center}
\includegraphics[
height=2.9611in,
width=3.5129in
]%
{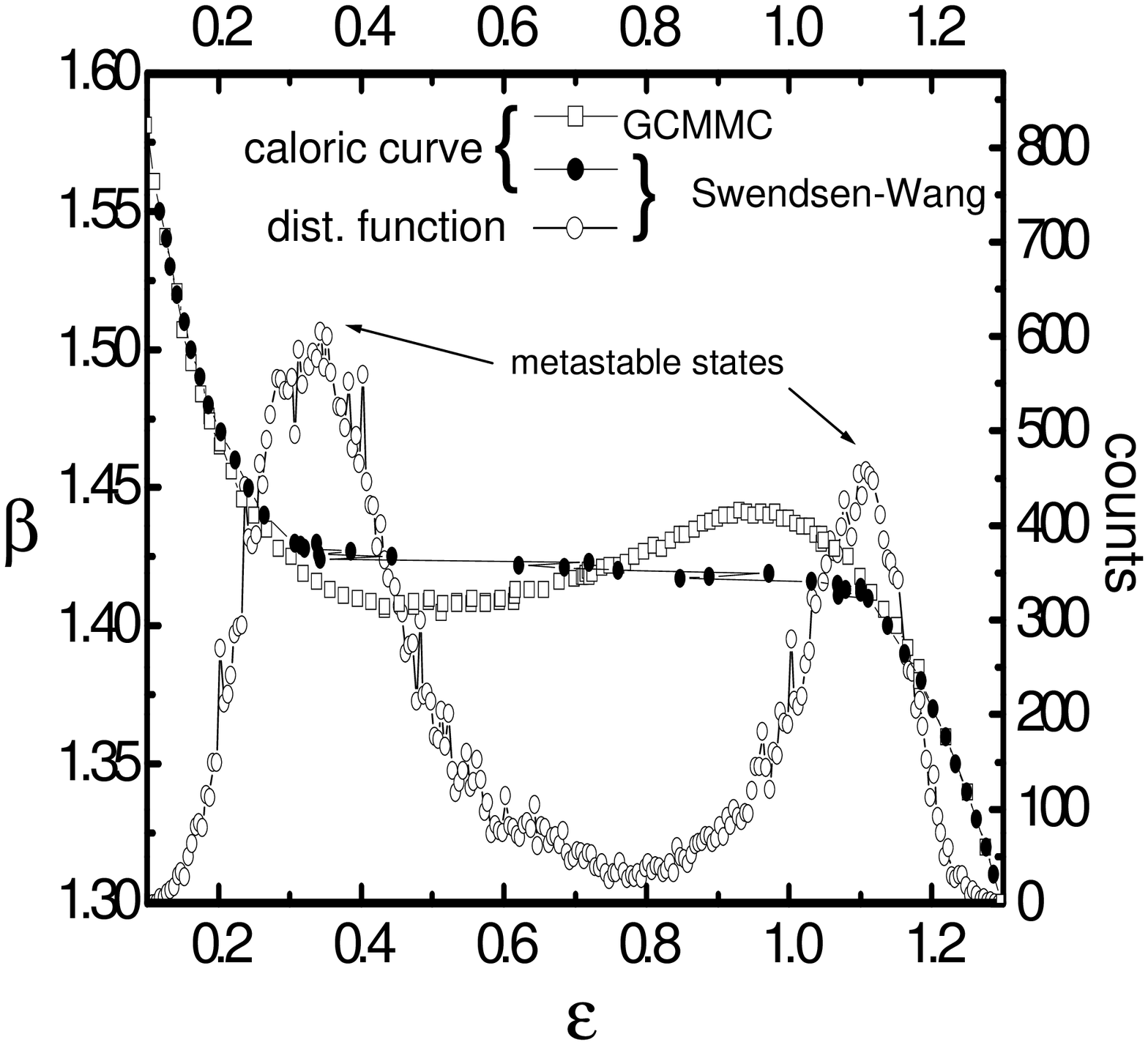}%
\caption{Comparative study between the GCMMC and the Swendsen-Wang (SW)
cluster algorithm. The SW algorithm is unable to describe the microcanonical
states with a negative heat capacity. Notice the bimodal character of the
energy distribution function at $\beta=1.42$. }%
\label{gc_sw.eps}%
\end{center}
\end{figure}

A preliminary calculation by using the MMC algorithm allows us to set a
anomalous region within the energetic windows $\left(  \varepsilon
_{1},\varepsilon_{2}\right)  $ with $\varepsilon_{1}=0.2$ and $\varepsilon
_{2}=1.2$ in which takes place the critical slowing down associated with the
existence of a first-order phase transition in this model for $L=25$. The
inverse critical temperature was estimated as $\beta_{c}\simeq1.4$, allowing
us to set $\lambda=0.8$. The main microcanonical observables obtained by using
the GCMMC algorithm with $n=10^{5}$ Metropolis iterations for each points is
shown in the FIG.\ref{thermo1.eps}: the entropy per particle $s\left(
\varepsilon\right)  $, the caloric $\beta\left(  \varepsilon\right)  =\partial
s\left(  \varepsilon\right)  /\partial\varepsilon$ and \textit{curvature}
$\kappa\left(  \varepsilon\right)  =\partial^{2}s\left(  \varepsilon\right)
/\partial\varepsilon^{2}$ curves. A comparative study between the present
method and the Swendsen-Wang clusters algorithm \cite{wang2} is shown in the
FIG.\ref{gc_sw.eps}.%

%

Notice that the heat capacity $c\left(  \varepsilon\right)  =-\beta^{2}\left(
\varepsilon\right)  /\kappa\left(  \varepsilon\right)  $ becomes negative when
$\varepsilon\in\left(  \varepsilon_{a},\varepsilon_{b}\right)  $ with
$\varepsilon_{a}\simeq0.51$ and $\varepsilon_{b}\simeq0.93$, which is a
feature of a first-order phase transition in a small system becoming extensive
in the thermodynamic limit. Such anomalous behavior is directly related with
the backbending in the caloric curve $\beta$ \textit{versus} $\varepsilon$ and
the existence of a convex intruder in the relative microcanonical entropy per
particle $\Delta s$ \textit{versus} $\varepsilon$ shown in the inserted graph.
This last dependence was obtained from a simple numerical integration of the
caloric $\beta\left(  \varepsilon\right)  $ and curvature $\kappa\left(
\varepsilon\right)  $ dependences by using the scheme:
\begin{align}
\delta s\left(  \varepsilon\right)   &  =s\left(  \varepsilon+\delta
\varepsilon\right)  -s\left(  \varepsilon\right)  ,\nonumber\\
&  \simeq\beta\left(  \varepsilon\right)  \delta\varepsilon+%
\frac12
\kappa\left(  \varepsilon\right)  \delta\varepsilon^{2},
\end{align}
based on the second-order approximation of the power expansion of the entropy.
We clarify to the reader that the true dependence plotted in this figure is
given by $\Delta s^{\ast}\left(  \varepsilon\right)  =s^{\ast}\left(
\varepsilon\right)  -s^{\ast}\left(  \varepsilon_{\ast}\right)  $, where
$s^{\ast}\left(  \varepsilon\right)  =s\left(  \varepsilon\right)
-\alpha\varepsilon$, with $\varepsilon_{\ast}=0.015$ and $\alpha=1.41$, in
order to make more evidence the existence of the convex intruder. It can be
shown that the convex intruder disappears progressively with the increasing of
the system size until becoming the Maxwell line represented in the inserted
graph of the FIG.\ref{thermo1.eps}, which provides us information about the
transition temperature and the latent heat of the first-order phase transition
undergone by this model system.

Since the heat capacity is always positive within the canonical ensemble, such
anomalous regions are inaccessible in this description. This fact evidences
the existence of a significant \textit{lost of information }about the
thermodynamical features\ of the system during the occurrence of a first-order
phase transitions when the canonical ensemble is used instead of the
microcanonical one. This difficulty is successfully overcome by the GCMMC
algorithm, which is able to predict the microcanonical average of the
microscopic observables in these anomalous regions where any others Monte
Carlo methods based on the consideration of the Gibbs canonical ensemble such
as the original MMC, the Swendsen-Wang and the Wolff single cluster algorithms
certainly do not work.

This fact is clearly illustrated in the FIG.\ref{gc_sw.eps}, which evidences
that the Swendsen-Wang algorithm is unable to describe the thermodynamic
states with a negative heat capacity: its results within the anomalous region
differ significantly from the ones obtained by using the GCMMC algorithm. The
Swendsen-Wang dynamics exhibits here an erratic behavior originated from the
competition of the two metastable states present in the neighborhood of the
critical point (the energy distribution function in the canonical ensemble is
bimodal when $\beta\in\left(  \beta_{1},\beta_{2}\right)  $, where $\beta
_{1}=1.405$ and $\beta_{2}=1.445$, being this feature the origin of the
supercritical slowing down).

As already shown by Gross in ref.\cite{gro na}, a convex intruder can be
associated with the existence of a \textit{non-vanishing interphase surface
tension} during the first-order phase transitions in systems with short-range
interactions outside the thermodynamic limit. Nevertheless, we will show in
the next subsection that this is not the only one information which could be
hidden behind the presence of a negative heat capacity.%

\begin{figure}
[ptb]
\begin{center}
\includegraphics[
height=3.3122in,
width=3.3641in
]%
{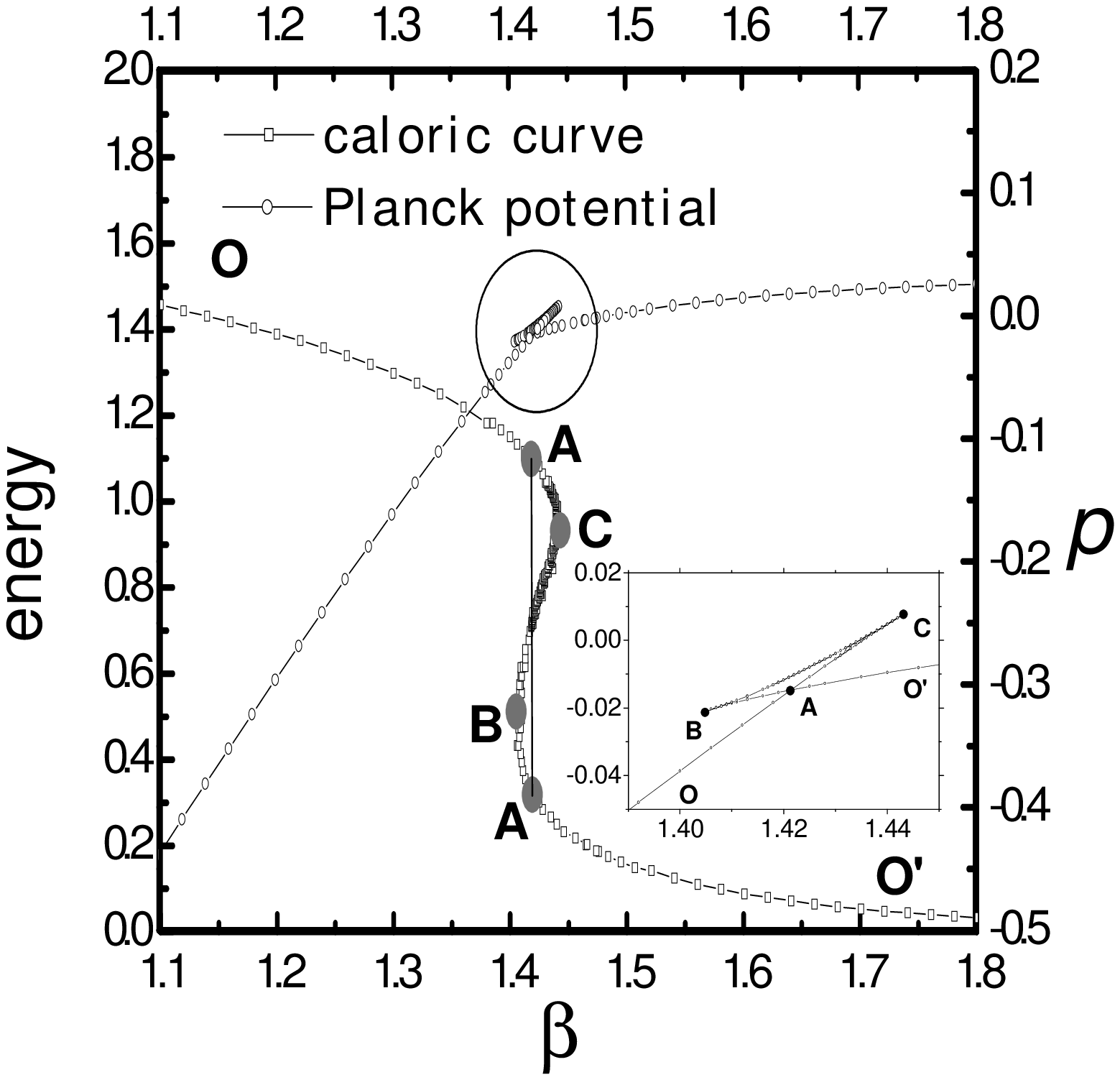}%
\caption{Canonical description of the $q=10$ states Potts model on
$25\times25$ square lattice in which are represented the caloric curve
$\varepsilon$ \textit{versus }$\beta$ and the Planck thermodynamic potential
per particles $P$ \textit{versus} $\beta$ dependences. The inserted graph is a
magnification of the region enclosed in the circle indicating the exact point
of the first-order phase transition. }%
\label{thermo2.eps}%
\end{center}
\end{figure}

We show in FIG.\ref{thermo2.eps} the canonical description of this model: the
caloric curve and the microcanonical Planck thermodynamic potential per
particle $p_{m}\left(  \varepsilon\right)  $\ \textit{versus} $\beta\left(
\varepsilon\right)  $ obtained from the Legendre transformation $p_{m}\left(
\varepsilon\right)  =\beta\left(  \varepsilon\right)  \varepsilon-s\left(
\varepsilon\right)  $. The branches \textbf{OA} and \textbf{AO'} represent the
canonically stable thermodynamic states corresponding to the minimal values of
the microcanonical Planck potential at a given $\beta$ (the true Planck
thermodynamic potential per particle of the canonical description when the
system size $N$ is large enough can be approximated by:
\begin{equation}
p_{c}\left(  \beta\right)  =-\ln Z\left(  \beta,N\right)  /N\simeq\min
_{\beta\left(  \varepsilon\right)  =\beta}\left\{  p_{m}\left(  \varepsilon
\right)  \right\}  , \label{cpp}%
\end{equation}
being $Z\left(  \beta,N\right)  $ the canonical partition function). Thus, the
points \textbf{A} represent the critical point of the first-order phase
transitions where $\beta_{A}\simeq1.421$, which is recognized from the
condition:
\begin{equation}
\text{ }\beta_{A}=\beta\left(  \varepsilon_{A}^{\left(  -\right)  }\right)
=\beta\left(  \varepsilon_{A}^{\left(  +\right)  }\right)  \text{ and
}p\left(  \varepsilon_{A}^{\left(  -\right)  }\right)  =p\left(
\varepsilon_{A}^{\left(  +\right)  }\right)  ,
\end{equation}
and appreciated in the inserted graph. Obviously, the first derivative of the
canonical Planck thermodynamic potential \ref{cpp} exhibits a discontinuity at
$\beta=\beta_{A}$. Since $\varepsilon_{A}^{\left(  -\right)  }\simeq1.099$ and
$\varepsilon_{A}^{\left(  +\right)  }\simeq0.319$, the estimated latent heat
$q_{lh}$ associated to this phase-transition is given by $q_{lh}%
=\varepsilon_{A}^{\left(  -\right)  }-\varepsilon_{A}^{\left(  +\right)
}\simeq0.78$.

The energy distribution function within the canonical ensemble is bimodal
inside the interval $\left(  \beta_{C},\beta_{B}\right)  $ with $\beta
_{B}\simeq1.405$ and $\beta_{C}\simeq1.443$, which is directly related with
the existence of metastable states, the branches \textbf{AC} (supercooled
disordered states)\ and \textbf{AB} (superheated ordered states), which are
the origin of the supercritical slowing down behavior observed during the
Monte Carlo simulations by using the ordinary Metropolis importance sample
algorithm MMC (the exponential divergence of the correlation time with the
system size increasing).

The branch \textbf{BC} is canonically unstable since its points represent
thermodynamical states with a negative heat capacity associated to the convex
intruder of the microcanonical entropy, where $\varepsilon_{B}\simeq0.933$ and
$\varepsilon_{C}\simeq0.512$. The energy region $\left(  \varepsilon
_{C},\varepsilon_{B}\right)  $ is practically invisible within the canonical
description when the system is large enough, and consequently, the ordinary
Metropolis algorithm based on the Gibbs ensemble will never access there as a
consequence of the ensemble inequivalence. Fortunately, the Metropolis
algorithm based on the reparametrization invariance (the using of generalized
canonical ensembles) considered in the present study overcomes successfully
all those the difficulties undergone by the original method

\subsection{Magnetic properties for $q=10$\label{magnet}}

The Potts model can be easily rephrased as ferromagnetic system with the
introduction of the bidimensional vector variable $\mathbf{s}_{i}=\left[
\cos\left(  \kappa\sigma_{i}\right)  ,\sin\left(  \kappa\sigma_{i}\right)
\right]  $ with $\kappa=2\pi/q$, where the total magnetization is simply given
by $\mathbf{M}=\sum_{i}\mathbf{s}_{i}$. Their magnetic thermostatistical
properties will be\ studied in the present subsection by means of the\ average
magnetization $\left\langle \mathbf{M}\right\rangle $ and the square
dispersion $G=\left\langle \left(  \mathbf{M}-\left\langle \mathbf{M}%
\right\rangle \right)  ^{2}\right\rangle $ which characterizes the
fluctuations of this microscopic observable.

It is well-known that the square dispersion of the total magnetization $G$
provides us a measure $g$ about the \ \textit{spatial correlations} between
the spin particles%
\begin{equation}
g=\frac{G}{N}\equiv\frac{1}{N}\sum_{ij}g_{ij},
\end{equation}
where $g_{ij}=\left\langle \mathbf{s}_{i}\cdot\mathbf{s}_{j}\right\rangle
-\left\langle \mathbf{s}_{i}\right\rangle \cdot\left\langle \mathbf{s}%
_{j}\right\rangle $ is the two-point correlation function. This quantity is
also related with the magnetic susceptivity $\chi\equiv dM/dB$ of the system
under the presence of an external magnetic field $\mathbf{B}$ (coupled to the
total magnetization modifying the original Hamiltonian $H$ \ref{definition} as
follows $H_{B}=H-\mathbf{B}\cdot\mathbf{M}$) throughout the called
\textit{static susceptibility sum rule} \cite{Gold}:%
\begin{equation}
\frac{\partial P}{\partial B}=\beta M,~\frac{\partial M}{\partial B}=\beta
G_{c},
\end{equation}
within the canonical description, while the corresponding identities in the
microcanonical ensemble are given by:
\begin{equation}
\frac{\partial S}{\partial B}=\beta M,~\frac{\partial M}{\partial B}=\beta
G_{m}+M\frac{\partial M}{\partial E}, \label{susceptibity}%
\end{equation}
where $P$ and $S_{B}$ are the Planck potential of the canonical ensemble and
the Boltzmann entropy of the microcanonical ensemble respectively,
$B\equiv\left\vert \mathbf{B}\right\vert $ and $M$ is the projection of the
total magnetization along the external magnetic field vector (see
demonstration in appendix \ref{others}).

It is very important to remark that although the existence of the ensemble
equivalence in the thermodynamic limit allows us to identify asymptotically
the expectation values of the microscopic observables (like the total
magnetization $\mathbf{M}$), those quantities characterizing their
fluctuations ($G$) and the response functions ($\chi$) depend essentially on
the nature of the statistical ensemble used in the description. For example,
while the canonical susceptibility is always positive, $\chi_{c}=\beta
G_{c}\geq0$, its corresponding microcanonical quantity $\chi_{m}$ could be
negative as a consequence of the presence of the term $M\partial M/\partial E$
in the second relation of the equation \ref{susceptibity} since magnetization
decreases in the ferromagnetic systems with the energy increasing.

Let us return to the study of the magnetic properties of the Potts model. It
is very easy to see that the Hamiltonian \ref{definition} is invariant under
the discrete group of transformations $\Pi_{q}$ composed by the $q!$
permutations among $q$ spin states. The subgroup of\ cyclic permutations
$Z_{q}=\left\{  T_{k};k=1,2,...q\right\}  $ is also a subgroup of the group of
bidimensional rotations $U\left(  2\right)  $, whose \textit{k-th}
transformation $T_{k}$\ acting on a given vector variable $\mathbf{s}\left(
\sigma\right)  =\left[  \cos\left(  \kappa\sigma\right)  ,\sin\left(
\kappa\sigma\right)  \right]  $ induces a rotation in an angle $\Delta
\varphi=2\pi k/q$:
\begin{equation}
T_{k}\mathbf{s}\left(  \sigma\right)  =\mathbf{s}\left(  \sigma+k\right)  .
\end{equation}
Obviously, the existence of this last symmetry leads\ trivially to the
vanishing of the exact statistical average of the magnetization.

However, a net magnetization can appears at low energies during the
paramagnetic-ferromagnetic phase transition (from disordered states towards
the ordered ones) as a consequence of the spontaneous symmetry breaking
associated to the nontrivial occurrence of an \textit{ergodicity breaking }in
the underlying dynamical behavior of this model system. This dynamical
phenomenon manifests itself when the time averages and the ensemble averages
of certain macroscopic observables can not be identified due to the
microscopic dynamics is effectively trapped in different subsets of the
configurational or phase space during the imposition of the thermodynamic
limit $N\rightarrow\infty$ \cite{Gold}.%

\begin{figure}
[ptb]
\begin{center}
\includegraphics[
height=2.6515in,
width=3.2396in
]%
{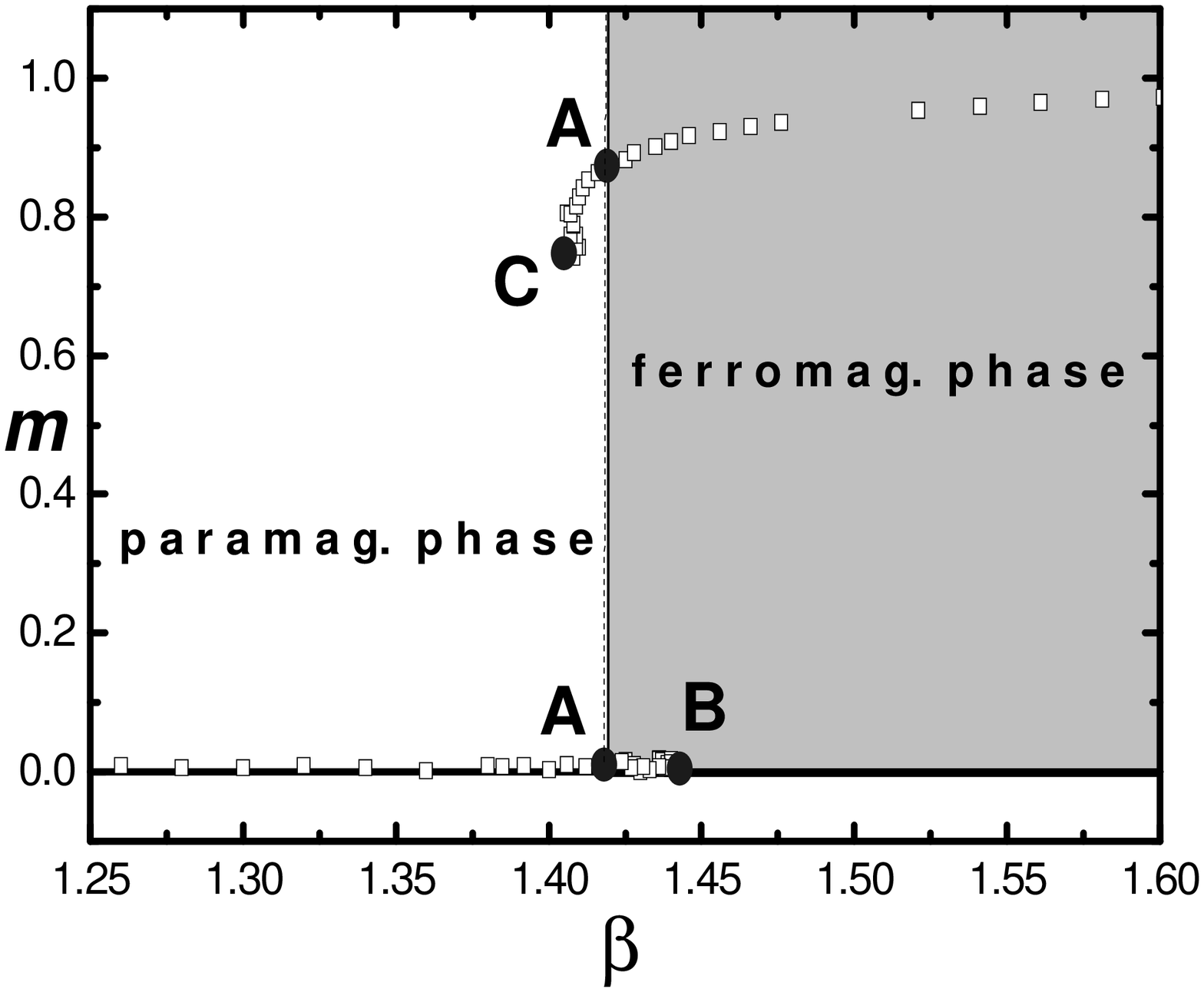}%
\caption{Magnetization $m$ versus $\beta$ dependence o the $q=10$ states Potts
model with $L=25$, which shows the existence of a discontinuous
ferromagnetic-paramagnetic phase transition at $\beta=\beta_{A}=1.421$.}%
\label{magnet.cal.eps}%
\end{center}
\end{figure}

We show in the FIG.\ref{magnet.cal.eps} the magnetization density
$m=\left\vert \left\langle \mathbf{M}\right\rangle \right\vert /N$ within the
canonical ensemble, where the stable (\textbf{OA} and \textbf{AO'}) and
metastable branches (\textbf{AC} and \textbf{AB}) are clearly visible. The
existence of a ferromagnetic-paramagnetic phase transition at the critical
point $\beta_{A}$ shows clearly that the $Z_{q}$ symmetry has been
spontaneously broken. However, it is remarkable how a net magnetization
appears \textit{abruptly}, that is, with a discontinuous character, a behavior
which is very similar to the one observed during the solid-liquid first-order
phase transition where the traslational symmetry is also spontaneously broken.%

\begin{figure}
[ptb]
\begin{center}
\includegraphics[
height=2.911in,
width=3.2396in
]%
{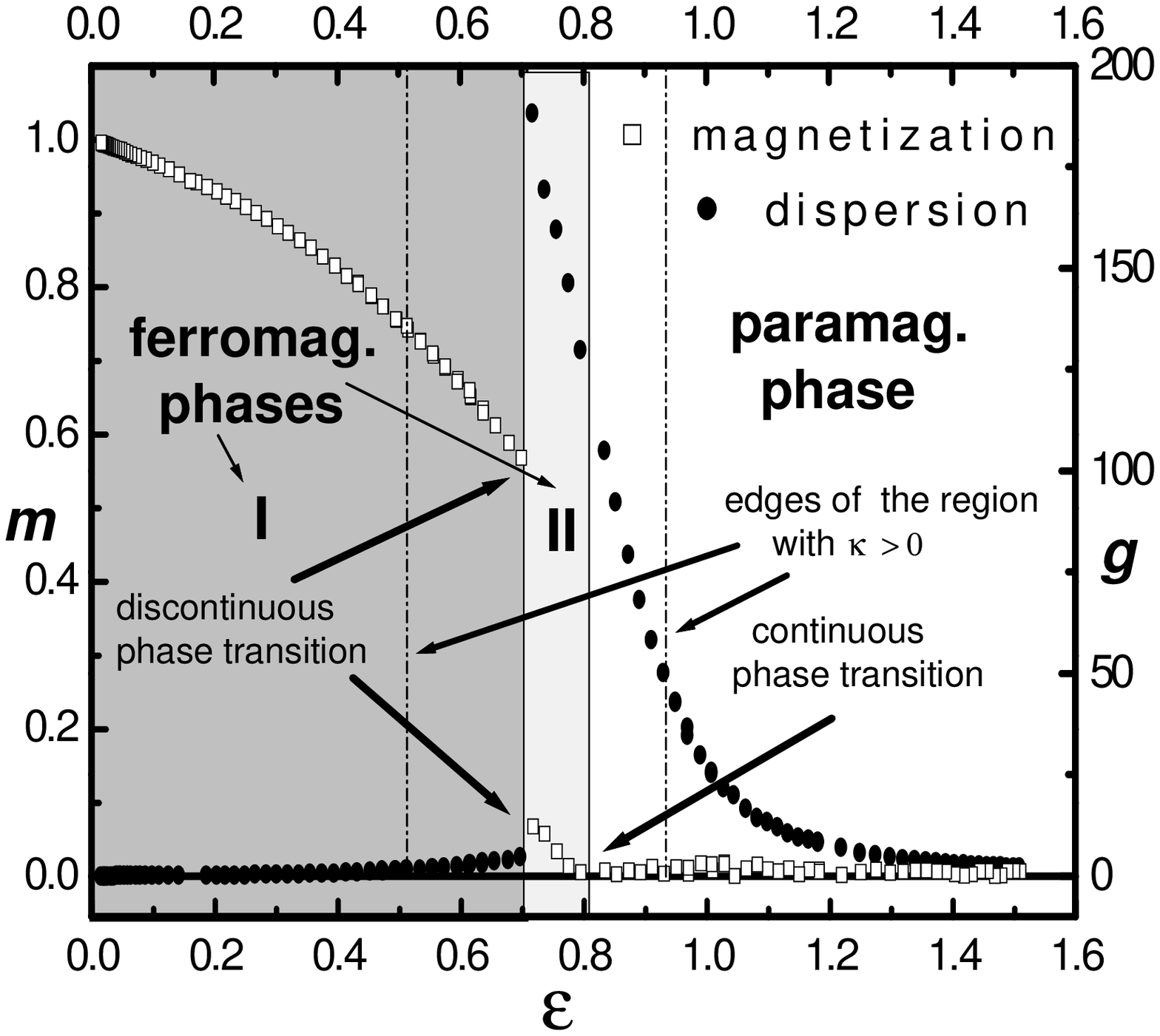}%
\caption{Magnetic properties of the model within the microcanonical ensemble
in which is shown the existence of two phase transitions at $\varepsilon
_{ff}\simeq0.7$ (ferro-ferro) and $\varepsilon_{fp}\simeq0.8$ (ferro-para).}%
\label{magnet1.eps}%
\end{center}
\end{figure}

As already commented\ in the above subsection, there are many thermodynamic
information which could be hidden behind a negative heat capacity during the
canonical description. Particularly, the reader may agree with us that it is
reasonable to expect within the microcanonical description the existence of a
critical energy where ferromagnetic-paramagnetic phase transition takes place
without the discontinuous character of the magnetization curve observed within
the canonical description. Surprisingly, reader can notice that
the\ qualitative behavior of the magnetic properties of this model system
shown in the FIG.\ref{magnet1.eps} (magnetization density $m$ and the
dispersion $g=G/N$) are much more interesting than our preliminary idea.

\begin{figure}
[ptb]
\begin{center}
\includegraphics[
height=2.559in,
width=3.2396in
]%
{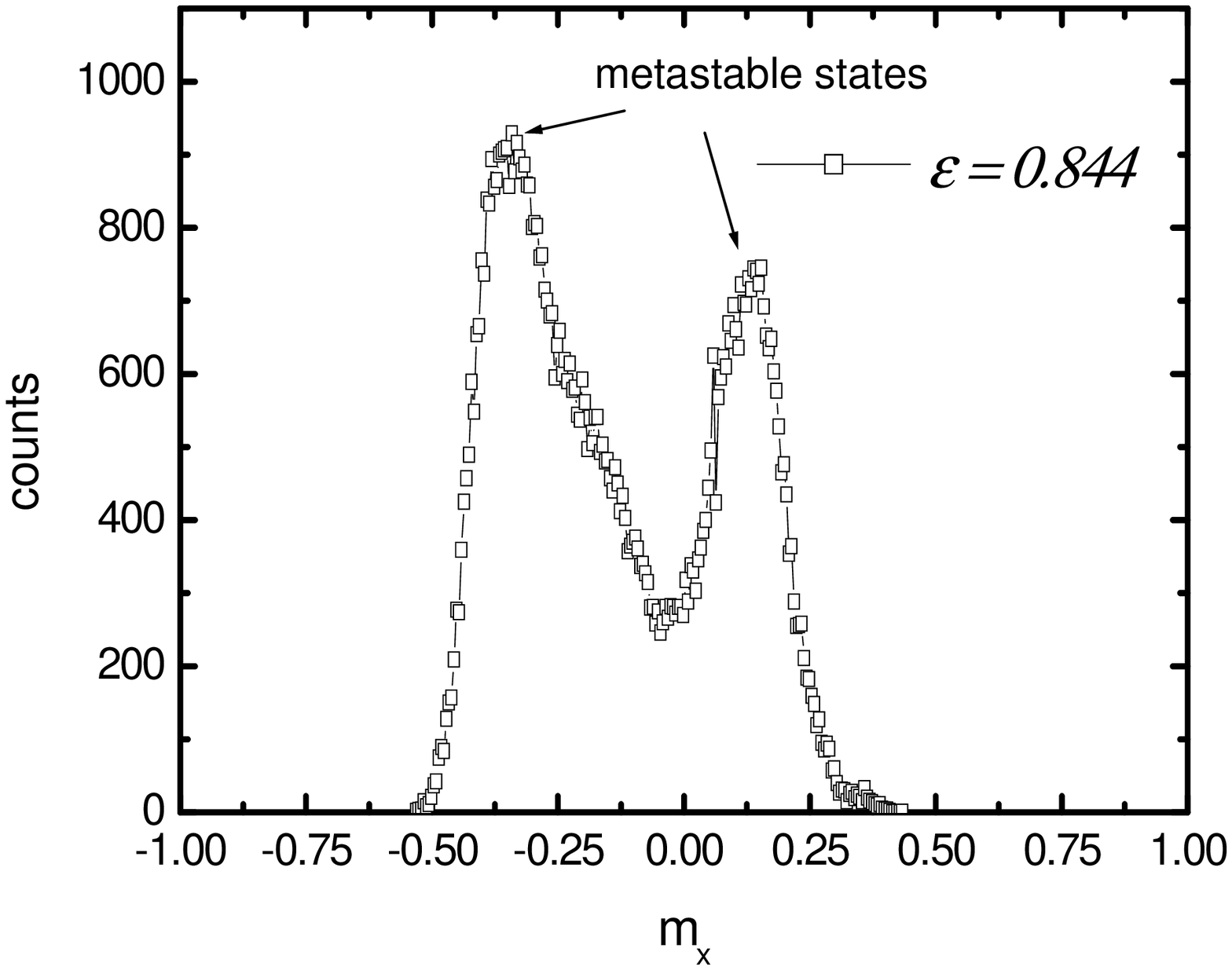}%
\caption{Histogram of the projection of the microscopic magnetization along
the direction of the spontaneous magnetization of the ferromanet type I, which
demostrates the existence of metastable states when $\varepsilon=0.844$.}%
\label{metastable.eps}%
\end{center}
\end{figure}

This magnetization density $m$ \textit{versus} $\varepsilon$ dependence
evidence clearly what could be considered as two phase transitions within the
microcanonical description of the $q=10$ states Potts model for $L=25$: a
\textit{continuous} (para-ferro) phase transition at the critical point
$\varepsilon_{fp}\simeq0.8$, and a \textit{discontinuous} (ferro-ferro)\ phase
transition at $\varepsilon_{ff}\simeq0.7$. Most of thermodynamic points were
obtained from a data of $n=10^{5}$ Metropolis iterations, with the exception
of all those points belonging to the energetic interval $\left(
0.7,0.93\right)  $ where $n=5\times10^{6}$ iterations were needed in order to
reduce the significant dispersion of the expectation values close to the
critical points. The large dispersions observed throughout the dispersion $g$,
the large relaxation times during the Metropolis dynamics, and the qualitative
behavior of the magnetization density strongly suggest us the presence of
several metastable states with different magnetization states at a given
energy within this last region, which is demonstrated in the
FIG.\ref{metastable.eps}.%

The microcanonical continuous phase transition between the
paramagnetic-ferromagnetic type I \ (with low magnetization density) phases
takes place with the spontaneous breaking of the $Z_{q}$ symmetry associated
to the occurrence of an ergodicity breaking in the microscopic dynamics.
Apparently, the large fluctuations and long-range correlations ordinarily
associated to this kind of phase transition are overlapped with the large
fluctuations associated to the existence of metastable states.

On the other hand, the $Z_{q}$ symmetry has been already spontaneously broken
during the occurrence of the microcanonical discontinuous phase transition
between the ferromagnet type I - ferromagnet type II \ (with high
magnetization density). However, this phase transition is also associated to
the occurrence of an ergodicity breaking in the underlying microscopic
picture: the microscopic dynamics of this model system can be effectively
trapped in the thermodynamic limit $L\rightarrow\infty$ in any of the
metastable states with different magnetization density present in the
neighborhood of this critical point.%

\begin{figure}
[ptb]
\begin{center}
\includegraphics[
height=2.9153in,
width=3.2396in
]%
{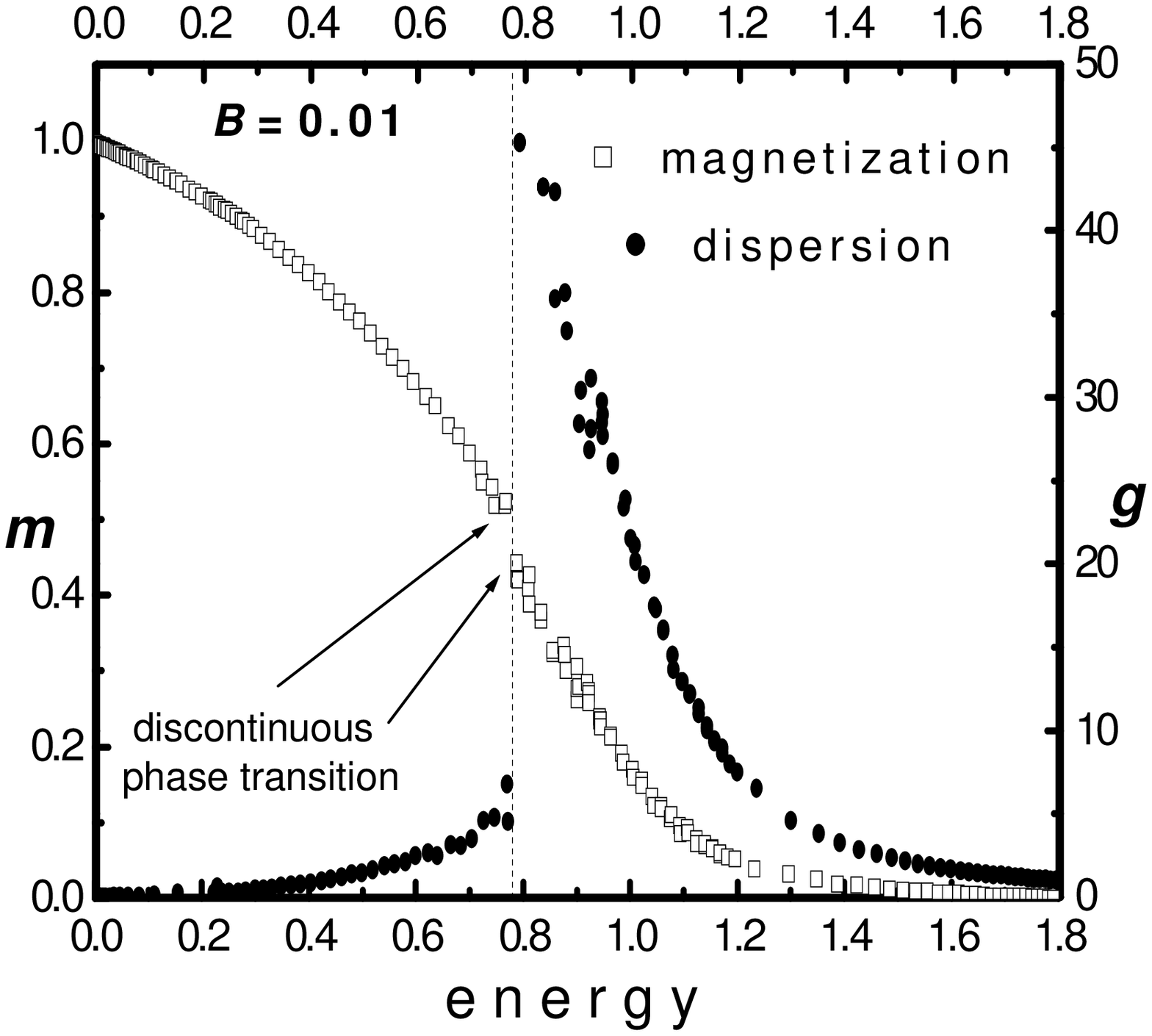}%
\caption{Magnetic properties of the model under the presence of an external
magnetic field with $B=0.01$. Notice that in this case the dispersion is not
large than in the first case with $B=0$, but it is clearly visible the
existence of the ferromagnetic-paramagnetic phase transition.}%
\label{magnet2.eps}%
\end{center}
\end{figure}
%

\begin{figure}
[h]
\begin{center}
\includegraphics[
height=2.936in,
width=3.2396in
]%
{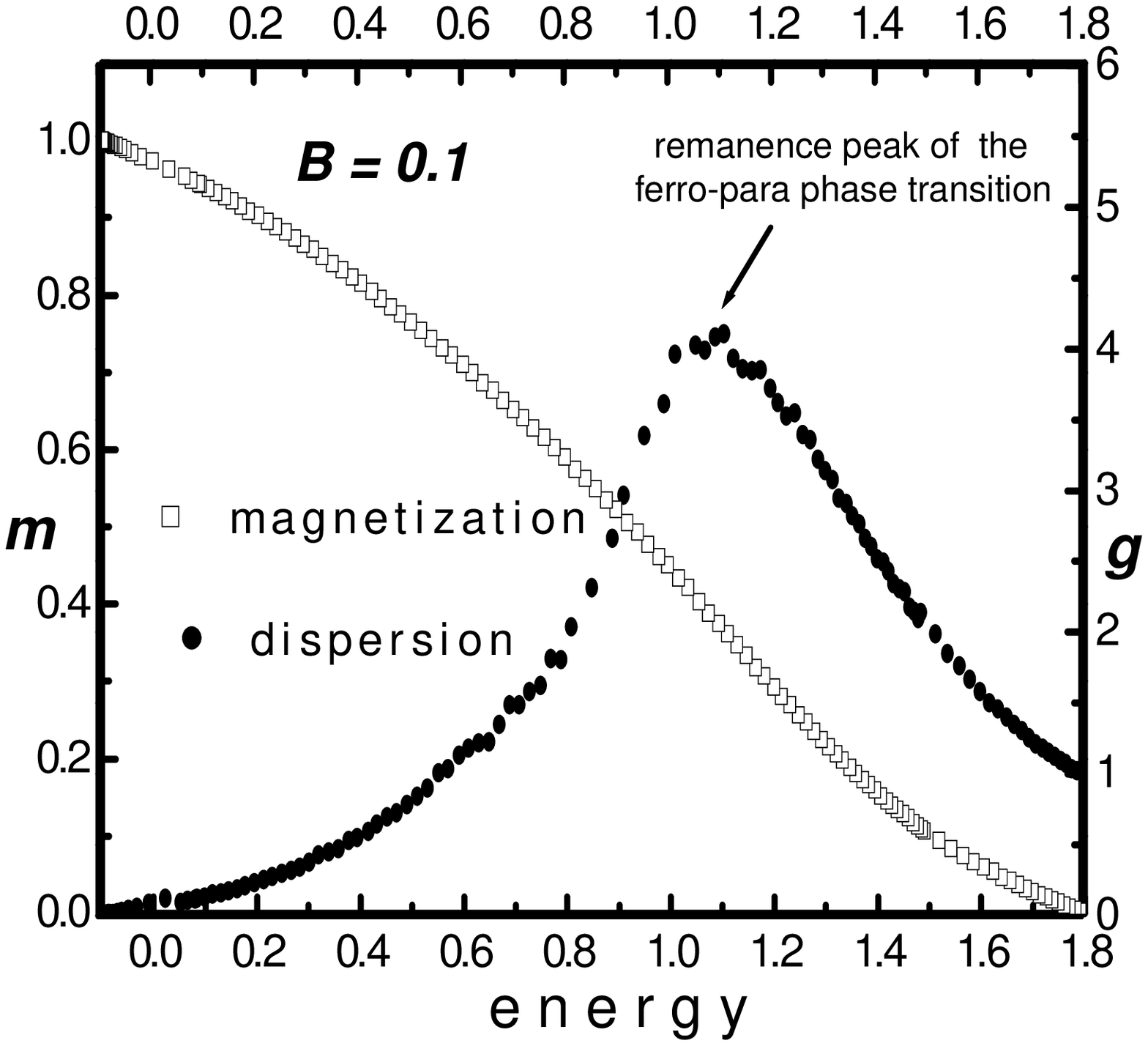}%
\caption{The same dependences of the FIGs (\ref{magnet1.eps}) and
(\ref{magnet2.eps}), but now under the influence of an external magnetig field
with $B=0.1$. }%
\label{magnet3.eps}%
\end{center}
\end{figure}

The FIGs \ref{magnet2.eps} and \ref{magnet3.eps} show the magnetic properties
of the model system under the presence of an external magnetic field with
$B=0.01$ and $B=0.1$. The reader can notice in the FIG.\ref{magnet2.eps} that
the microcanonical discontinuous phase transition is also present for low
intensities of the external magnetic field, but the discontinuity observed in
the magnetization $m\left(  \varepsilon\right)  $\ and the dispersion
$g\left(  \varepsilon\right)  $ dependences is not so abrupt in this case.
Apparently, such behavior is reduced progressively with the $B$ increasing
until disappear when the magnetic field intensity is large enough, as already
illustrated in the FIG.\ref{magnet3.eps}. This last figure shows that the
dispersion curve has a peak, which is a remanence of the
paramagnetic-ferromagnetic phase transition when $B\not =0$. Reader may notice
that the presence of discontinuous phase transition affects significantly the
qualitative behavior of the square dispersion $g$ of the magnetization.

\subsection{What can be learned from the microcanonical description of this
model system?}

The backbending behavior in the microcanonical caloric curve shown in the
FIG.\ref{thermo1.eps} (becoming a plateau in the thermodynamic limit) is
usually interpreted as the signature of a first-order phase transition (with
spontaneous symmetry breaking). On the other hand, the magnetization curve
shown in the FIG.\ref{magnet1.eps} exhibits two anomalies which can be also
considered as phase transition of the thermostatistical description of this
model system: a continuous phase transition (with spontaneous symmetry
breaking) and a discontinuous phase transition (where the underlying symmetry
has been already broken). Consequently, what kind of phase transition exhibits
the thermostatistical description of the $q=10$ states Potts model? The answer
of the above question in our viewpoint \textit{depends on the external
conditions imposed to the system}.

Obviously, this model system put in contact with a Gibbs thermostat undergoes
a discontinuous first order phase transition while a smooth change of the
inverse temperature $\beta$ in the neighborhood of the critical point
$\beta_{A}=1.421$. There, the phase coexistence phenomenon characteristic of
this kind of phase transitions can be observed and the region $\left(
\varepsilon_{C},\varepsilon_{B}\right)  $ is inaccessible within the Gibbs
canonical description in the thermodynamic limit.

A different picture is revealed when the model system is isolated
(microcanonical description). All the anomalous region $\left(  \varepsilon
_{C},\varepsilon_{B}\right)  $ is now accessible. Many microcanonical states
there show no anomalous behavior with the exception of those macrostates
within the region $\left(  \varepsilon_{ff},\varepsilon_{B}\right)  $ which
are affected by the incidence of metastable states. The imposition of the
thermodynamic limit leads to the suppression of the metastable states, but the
existence of anomalous behaviors persists in the neighborhood of the
para-ferro continuous phase transition at $\varepsilon_{fp}$ and the
ferro-ferro discontinuous phase transition at $\varepsilon_{ff}$.
Interestingly, such anomalies can not be apparently associated with any
anomaly of the caloric or curvature curves in the FIG.\ref{thermo1.eps}, but
by using the microcanonical magnetization curve shown in FIG.\ref{magnet1.eps}.

Taking into account the microcanonical thermodynamic identities in the
thermodynamic limit $N\rightarrow\infty$ shown in the equation
\ref{susceptibity} between the entropy $S$, the magnetization $M$, the
external magnetic field $B$ and the square dispersion $G$: the discontinuous
phase transition at $\varepsilon_{ff}$ corresponds to a discontinuity of the
first derivative of the entropy per particle:%
\begin{equation}
\lim_{\varepsilon\rightarrow\varepsilon_{ff}^{+}}\lim_{B\rightarrow0}%
\frac{\partial s\left(  \varepsilon,B\right)  }{\partial B}\not =%
\lim_{\varepsilon\rightarrow\varepsilon_{ff}^{-}}\lim_{B\rightarrow0}%
\frac{\partial s\left(  \varepsilon,B\right)  }{\partial B},
\end{equation}
while the continuous phase transition at $\varepsilon_{fp}$ corresponds to a
discontinuity of the second derivative of the entropy:%
\begin{equation}
\lim_{\varepsilon\rightarrow\varepsilon_{fp}^{+}}\lim_{B\rightarrow0}%
\frac{\partial^{2}s\left(  \varepsilon,B\right)  }{\partial\varepsilon\partial
B}\not =\lim_{\varepsilon\rightarrow\varepsilon_{fp}^{-}}\lim_{B\rightarrow
0}\frac{\partial^{2}s\left(  \varepsilon,B\right)  }{\partial\varepsilon
\partial B}. \label{discon}%
\end{equation}
Consequently, \textit{the present microcanonical anomalies can be recognized
by the lost of analyticity of the entropy per particle in the thermodynamic
limit, and the same ones are related with the occurrence of an ergodicity
breaking in the microscopic picture of this model system.}

\section{Towards a new classification scheme\label{scheme}}

The analysis developed in the above two sections allows us to classify the
thermodynamic anomalies type \textbf{A} and type \textbf{B} introduced in the
beginning of the section \ref{some} as follows. The anomaly type \textbf{A}
are just anomalies observed in open systems (canonical description) which are
dependent on the external experimental conditions due to they originated from
the ensemble inequivalence. The anomaly type \textbf{B} are all those
anomalies which are always observed in an isolated system (microcanonical
description) and can \textit{potentially} appear under any external
experimental conditions.

\subsection{The anomalies type \textbf{A}}

Anomalies type \textbf{A} correspond what we usually called as first-order
phase transitions in conventional Thermodynamics, and they are only relevant
for an open system since their existence depends crucially on the nature of
the external conditions imposed to the interest system. We recognized them by
the existence of a latent heat for the phase transition associated to the
multimodal character (a signature of the phase coexistence) of the energy
distribution function within the (generalized) canonical description. In terms
of the thermodynamic potentials, an anomaly type \textbf{A} manifests as a
discontinuity in some of the first derivatives the generalized Planck
potential, which are directly related to the existence of convex regions of
the microcanonical entropy $S$ in a given reparametrization $\Theta$ dependent
on the external experimental setup (the using of a generalized thermostat).

An anomaly type \textbf{A} represents in this way the inability of the
(generalized) canonical description in describing all those macrostates which
can be accessed within the microcanonical description. As already shown, this
kind of anomaly can be avoided by using another experimental setup which
ensures the global equivalence of the generalized canonical description with
the microcanonical ensemble. The Monte Carlo method exposed in the subsection
\ref{gmmc} is precisely based\ on this idea.

The Swensen-Wang algorithm is a Monte Carlo cluster algorithm based on the
consideration of the Gibbs canonical ensemble, that is, it simulates the
thermodynamical equilibrium of the Potts model system put in contact with an
ordinary thermostat (heat bath). Since the Gibbs canonical description of the
$q=10$ states Potts model is not globally equivalent to its microcanonical
description, a sudden change in the energy per particle $\varepsilon$\ of this
model system is observed in the neighborhood of the critical inverse
temperature $\beta_{c}\simeq1.421$. On the other hand, the GCMMC algorithm
describes the thermodynamic equilibrium of a given system in contact with a
generalized thermostat. As already shown in FIG.\ref{gc_sw.eps}, this change
of the external conditions allows the system to access to all those
inaccessible regions within the Gibbs canonical ensemble, and therefore,
neither there exist lost of information nor any sudden change of the energy
per particle is observed now. Thus, the thermodynamical study of the $q=10$
states Potts model shows that the first-order phase transitions are
\textit{avoidable} thermodynamical anomalies.

\subsection{The anomalies type \textbf{B}}

The lost of analyticity of the entropy per particle in the thermodynamic
limit, the anomaly type \textbf{B}, is obviously the only one mathematical
anomaly of the microcanonical entropy which is reparametrization invariant.
This character explains why this kind of anomaly could potentially appear
under any external experimental setup (which determines the specific
reparametrization $\Theta$ used in the generalized canonical ensemble), and
consequently, it is the only thermodynamical anomaly microcanonically
relevant. We say "potentially" because of there exists the possibility that an
anomaly type \textbf{B} could be hidden by the lost of information associated
to the ensemble inequivalence, i.e.: the continuous (ferro-para) and the
discontinuous (ferro-ferro) phase transitions described in the
FIG.\ref{magnet1.eps} are hidden within the Gibbs canonical description.
\begin{figure}
[t]
\begin{center}
\includegraphics[
trim=0.000000in 0.000000in -0.029399in 0.000000in,
height=1.5056in,
width=2.1543in
]%
{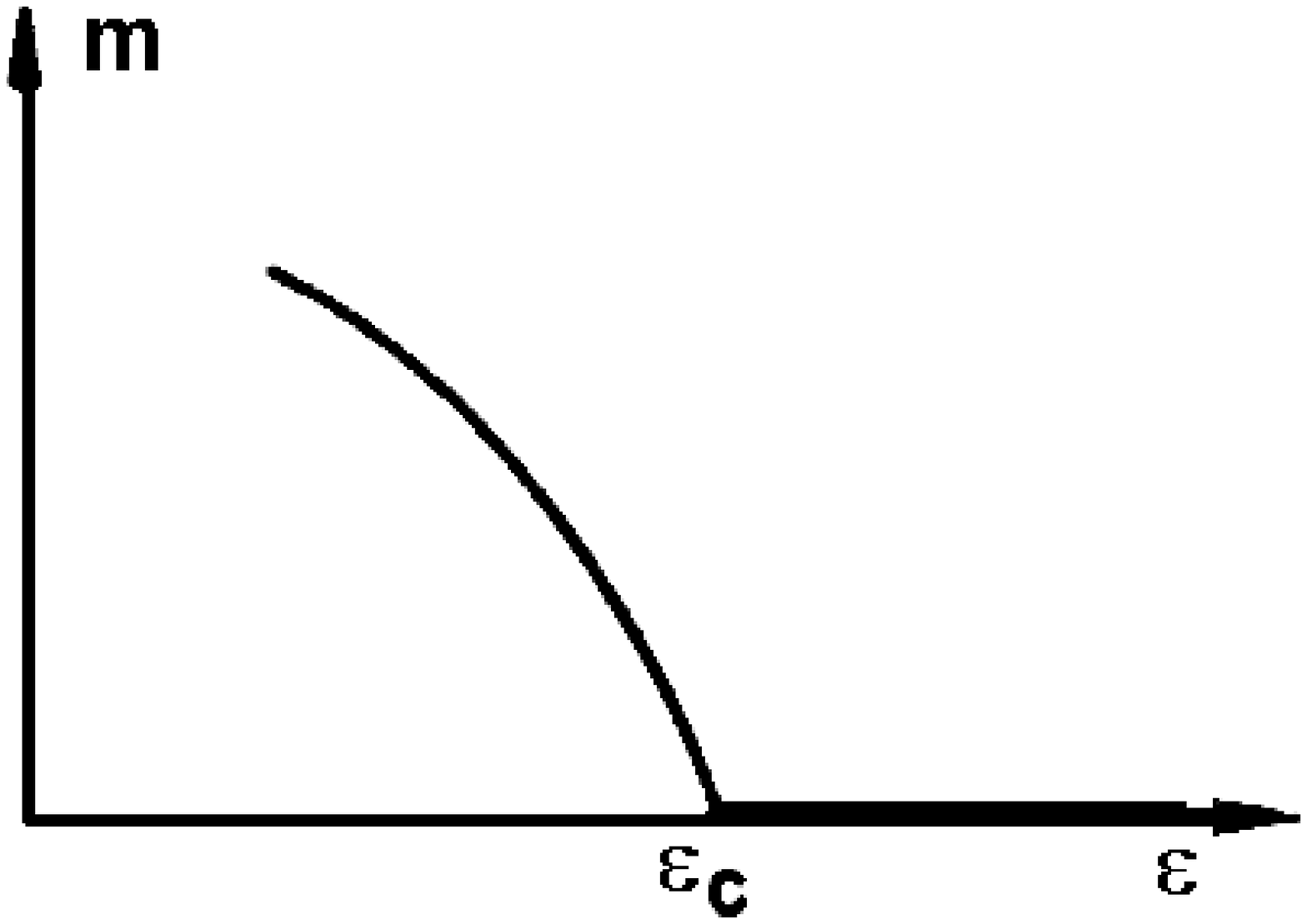}%
\caption{Schematic representation of the onset of the magnetization curve
within the microcanonical description in the limit of zero applied magnetic
field of a ferro-para phase transition.}%
\label{representa.eps}%
\end{center}
\end{figure}

Since the microcanonical ensemble is just a dynamical ensemble, the anomaly
type \textbf{B} should be the macroscopic manifestation of a sudden change in
the dynamical behavior of the isolated Hamiltonian system. As already pointed
in our study of the $q=10$ states Potts model, the lost of analyticity of the
entropy per particle in the thermodynamic limit seems to be related to the
\textit{ergodicity breaking phenomenon}. The well-known spontaneous symmetry
breaking associated to the ferro-para second order phase transitions
schematically represented in the FIG.\ref{representa.eps} is a generic example
of an ergodicity breaking which is always connected to a lost of analyticity
of the entropy per particle in the thermodynamic limit. The demonstration
starts from the consideration of the first identity of the equation
\ref{susceptibity}:
\begin{equation}
\frac{\partial s\left(  \varepsilon,B\right)  }{\partial B}=\beta\left(
\varepsilon,B\right)  m\left(  \varepsilon,B\right)  , \label{disc22}%
\end{equation}
which allows us to obtain the magnetization curve $m\left(  \varepsilon
,B\right)  $ from the entropy per particle in the thermodynamic limit
$s\left(  \varepsilon,B\right)  $, being $\beta\left(  \varepsilon,B\right)
=\partial s\left(  \varepsilon,B\right)  /\partial\varepsilon$ the
microcanonical caloric curve. The partial derivative $\partial/\partial
\varepsilon$ of the above equation is given by:
\begin{equation}
\frac{\partial^{2}s\left(  \varepsilon,B\right)  }{\partial\varepsilon\partial
B}=\frac{\partial\beta\left(  \varepsilon,B\right)  }{\partial\varepsilon
}m\left(  \varepsilon,B\right)  +\beta\left(  \varepsilon,B\right)
\frac{\partial m\left(  \varepsilon,B\right)  }{\partial\varepsilon}.
\end{equation}
While the first derivative of the caloric curve $\partial\beta\left(
\varepsilon,B\right)  /\partial\varepsilon$ always exists in short-range
interacting systems, the dependence $\partial m\left(  \varepsilon,B\right)
/\partial\varepsilon$ exhibits a discontinuity at the critical point
$\varepsilon_{c}$\ in the limit of zero applied magnetic field, and
consequently:%
\begin{equation}
h\left(  \varepsilon\right)  =\lim_{B\rightarrow0}\frac{\partial^{2}s\left(
\varepsilon,B\right)  }{\partial\varepsilon\partial B}%
\end{equation}
is discontinuous function.

The anomaly type \textbf{B} described above can be referred as a\textit{
microcanonical} \textit{continuous phase transition} since the first
derivatives of the entropy per particle in the thermodynamic limit are
continuous (\textit{Anomaly type B.I}). Contrary, we can referred a anomaly
type \textbf{B} as a \textit{microcanonical discontinuous phase transition}
when some of the first derivatives of the entropy per particle in the
thermodynamic limit are discontinuous at the point of lost of analyticity
(\textit{Anomaly type B.II}). Particularly, the ferro-ferro phase transition
observed in the microcanonical description of the $q=10$ states Potts model is
a clear example of a microcanonical discontinuous phase transition: Since the
magnetization curve is discontinuous, and therefore, the first derivative
\ref{disc22} is also discontinuous.

The reader can notice that most of the well-known continuous (second-order)
phase transitions in the conventional Thermodynamics corresponds to
microcanonical continuous phase transitions, since the applicability of the
Legendre transformation $P=\beta E-S$ during the ensemble equivalence in the
thermodynamic limit allows that every anomaly type \textbf{B.I} leads to lost
of analyticity of the Planck potential (or the Helmholtz free energy).
However, we will show below that all anomaly that could be classified as a
continuous phase transition within the conventional Thermodynamics does not
correspond necessarily to a microcanonical continuous phase transition.

A feature of the physical systems exhibiting a continuous phase transition
associated to the occurrence of an spontaneous symmetry breaking is the
existence of \textit{divergent power laws }behavior in the heat capacity (and
other response functions) in the neighborhood of the critical point $\beta
_{c}$:
\begin{equation}
c\left(  \beta\right)  =\left\{
\begin{array}
[c]{cc}%
A^{\left(  -\right)  }\left(  -t\right)  ^{-\alpha} & t<0,\\
A^{\left(  +\right)  }t^{-\alpha} & t>0,
\end{array}
\right.  \label{power}%
\end{equation}
being $t=\beta_{c}-\beta$, where the called critical exponent $\alpha$ and the
amplitude ratio $A^{\left(  +\right)  }/A^{\left(  -\right)  }$ are the same
within a universality class \cite{Gold}. The characteristic "$\lambda$-form"
of the continuous phase transitions is directly related to the presence of
divergent power laws with universal ratio $A^{\left(  +\right)  }/A^{\left(
-\right)  }\not =1$ and nontrivial critical exponent $\alpha$: i.e.:
$\alpha=\left(  0.110-0.116\right)  $ for the Ising universality class. The
divergence of the heat capacity in the continuous phase transitions can be
explained by an eventual vanishing of the second derivative of the entropy
$\partial^{2}s/\partial\varepsilon^{2}$ at the critical point. However, the
nature of the divergent power law depends crucially on the analyticity of the
entropy per particle in the thermodynamic limit.

Let be a hypothetical system whose entropy per particle in the thermodynamic
limit $s\left(  \varepsilon\right)  $ is analytical in a given region but its
second derivative $\partial^{2}s\left(  \varepsilon\right)  /\partial
\varepsilon^{2}$ vanishes eventually at a certain point $\varepsilon_{c}$ of
this region. This is just anomaly type \textbf{A} where the anomalous region
of the microcanonical entropy is composed by only one point. Since the caloric
curve $\beta\left(  \varepsilon\right)  =\partial s\left(  \varepsilon\right)
/\partial\varepsilon$ is bijective in this case, ensemble equivalence is
ensured. However, the average square dispersion of the system energy and the
heat capacity within the Gibbs canonical ensemble go to infinite at
$\varepsilon_{c}$, so that, the second derivative of the Planck thermodynamic
potential diverges at the corresponding critical point $\beta_{c}=\beta\left(
\varepsilon_{c}\right)  $. In spite of the present anomalous behavior can be
classified as a continuous phase transition in the conventional Thermodynamics
viewpoint, it does not correspond to any microcanonical phase transition.
Particularly, the missing of the lost of analyticity of the entropy allows us
to think that the above anomaly can not be associated to the occurrence of an
ergodicity breaking. For example, the analytical character of the entropy per
particle in the thermodynamic limit of the above hypothetical system allows us
to approximate the caloric curve $\beta\left(  \varepsilon\right)  $ in the
neighborhood of the critical point $\varepsilon_{c}$ by the Taylor power
series:
\begin{equation}
\beta=\beta_{c}-\lambda\left(  \varepsilon-\varepsilon_{c}\right)
^{2n+1}+...,
\end{equation}
where $\lambda>0$ and $n$ a positive integer, expression leading to a power
law divergent behavior of the heat capacity \ref{power} with critical exponent
$\alpha=2n/\left(  2n+1\right)  $ and $A^{\left(  -\right)  }=A^{\left(
+\right)  }=\beta_{0}^{2}/\lambda^{1-\alpha}\left(  2n+1\right)  $.
Consequently, this result corresponds to a critical phenomenon with universal
ratio $A^{\left(  +\right)  }/A^{\left(  -\right)  }\equiv1$ and critical
exponent $\alpha$ related to a odd number $2n+1=1/\left(  1-\alpha\right)  $.

The above result allows to claim that the existence of divergent power laws
with nontrivial critical exponents $\alpha$\ and universal ratio $A^{\left(
+\right)  }/A^{\left(  -\right)  }\not =1$ in most of real physical systems
exhibiting a continuous phase transition associated to the occurrence of a
spontaneous symmetry breaking is a clear indicator of the lost of analyticity
of the entropy per particle in the thermodynamic limit in such cases. An
anomaly like the one exhibited by our hypothetical system can be seen as an
asymptotic case of a discontinuous phase transition, which shall be referred
in this work as a \textit{1st-kind continuous phase transition}.

The discontinuous character of the first derivatives of the entropy per
particle in the thermodynamic limit during the occurrence of the
microcanonical discontinuous phase transitions can lead to discontinuity of
some of the first derivatives of the Planck potential (or other thermodynamic
potential characterizing an open system) or even provoke the discontinuity of
the thermodynamical potential itself. This last mathematical anomaly is
unusual for the systems dealt within the conventional Thermodynamics, but it
can be observed in the astrophysical context and others long-range interacting
systems. The interested reader can find in the ref.\cite{chava} an example of
a discontinuity of the caloric curve $\beta\left(  \varepsilon\right)  $
observed in the thermodynamical description of astrophysical model, indicating
the existence of metastable states at the same total energy with different
temperature (mathematical anomaly leading to a discontinuity of the Planck
potential $P=\beta E-S$). While the thermodynamical behavior associated to the
discontinuity of the Planck thermodynamic potential could be referred as a
\textit{zero-order phase transition} within the well-known Ehrenfest
classification, the discontinuity of some of its first derivatives is just a
discontinuous phase transition.\ 

Although we can not provide in this work a\ rigorous demonstration about the
relationship between the lost of analyticity of the entropy per particle in
the thermodynamic limit with the occurrence of ergodicity breaking in the
microscopic dynamics, we have considered in this work some examples suggesting
that such connection exists. Loosely speaking, the ergodicity breaking takes
place as a consequence of the dynamical competition among metastable states,
where the predominance of any of them depends crucially on the initial
conditions of the microscopic dynamics and the boundary conditions
\cite{gallavotti,Gold}. Particularly, the thermodynamical study of the $q=10$
states Potts model presented in the section \ref{potts} suggests that the
microcanonical continuous phase transition observed in this model system is
provoked by the competition among metastable states which essentially
identical (states with the same magnetization density) since they are related
by a symmetry transformation ($Z_{q}$). On the other hand, the microcanonical
discontinuous phase transition observed in this model system is also provoked
by the competition among metastable states, which are essentially different
(states with different spontaneous magnetization density) since them can not
be related by a symmetry transformation.

The reader may notice that the discontinuity of the first derivative of the
entropy described in the ref.\cite{chava} (a microcanonical discontinuous
phase transition) can be associated to the occurrence of ergodicity breaking
originated from the competition among metastable states with different temperature.

\subsection{Summary}

As already shown in this work, the existence and the features of anomalies
observed in the thermodynamical description of given Hamiltonian system
depends crucially on the external conditions which have been imposed. It means
that a phase transition is not only an intrinsic thermodynamical anomaly of a
given system, but also a specific response to the nature of the external
control of its thermodynamic equilibrium. Generally speaking, the mathematical
description of the phase transitions for systems in thermodynamic limit starts
from the consideration of the lost of analyticity of the thermodynamic
potential which is relevant in a given application.

The most simple characterization of the thermodynamical anomalies of a given
Hamiltonian system is obtained within the microcanonical description, which is
relevant when the interest system is isolated. Phase transitions here can be
recognized by the lost of analyticity of the entropy\ $S$ in the thermodynamic
limit, a mathematical anomaly (type \textbf{B.I} or \textbf{B.II}) which is
reparametrization invariant and should be originated from the occurrence of
the ergodicity breaking phenomenon. This results are summarized in the Table
\ref{isolated}.
\begin{table}[tbp] \centering
\begin{tabular}
[c]{|c|c|}\hline\hline
\textbf{Thermodynamical anomaly} & \textbf{Classification}\\\hline\hline
\multicolumn{1}{|l|}{$%
\begin{array}
[c]{l}%
\text{Type \textbf{A} (presence of non }\\
\text{concave regions of the entropy }S\text{)}\\
\text{(non reparametrization invariant)}%
\end{array}
$} &
\begin{tabular}
[c]{c}%
microcanonically\\
irrelevant
\end{tabular}
\\\hline
\multicolumn{1}{|l|}{$%
\begin{array}
[c]{c}%
\text{Type \textbf{B.I} (lost of analyticity of }\\
\multicolumn{1}{l}{\text{the entropy }S\text{\ with continuous}}\\
\multicolumn{1}{l}{\text{ first derivatives) }}\\
\multicolumn{1}{l}{\text{(reparametrization invariant)}}%
\end{array}
$} &
\begin{tabular}
[c]{l}%
microcanonical\\
continuous PT
\end{tabular}
\\\hline
\multicolumn{1}{|l|}{$%
\begin{array}
[c]{c}%
\text{Type \textbf{B.II} (lost of analyticity of }\\
\multicolumn{1}{l}{\text{the entropy }S\text{ with some}}\\
\multicolumn{1}{l}{\text{discontinuous first derivatives)}}\\
\multicolumn{1}{l}{\text{(reparametrization invariant)}}%
\end{array}
$} & \multicolumn{1}{|l|}{%
\begin{tabular}
[c]{c}%
microcanonical\\
discontinuous PT
\end{tabular}
}\\\hline\hline
\end{tabular}
\caption{Phase transitions (PT) within the microcanonical description
(isolated system). \label{isolated}}%
\end{table}%

The presence of the experimental setup controlling the thermodynamic
equilibrium of the interest system leads to a significant complexation of the
phase transitions. Phase transitions here are recognized by the lost of
analyticity of the relevant thermodynamic potential $\mathcal{P}$ (Planck or
Helmholtz free energy, gran potential or any other admissible generalization).

We have now \textit{five thermodynamical anomalies} for the open systems in
contract to the only \textit{two relevant} when they are isolated. Besides the
introduction of the zero-order phase transitions, we consider also a
distinction among those continuous and discontinuous phase transitions which
are related with a lost of analyticity of the entropy $S$. A tentative
classification scheme is summarized in the Table \ref{openned}.%

\begin{table}[tbp] \centering
\begin{tabular}
[c]{|c|c|}\hline\hline
\textbf{Thermodynamical anomaly} & \textbf{Classification}\\\hline\hline
\multicolumn{1}{|l|}{$%
\begin{array}
[c]{l}%
\mathcal{P}\text{ is discontinuous as consequence}\\
\text{of a discontinuity of control parame }\\
\text{ter }\eta=\partial S/\partial I\text{ used in the Legendre }\\
\text{transformation }\mathcal{P}=\eta I-S\text{.}%
\end{array}
$} & zero-order PT\\\hline
\multicolumn{1}{|l|}{$%
\begin{array}
[c]{l}%
\text{Some of first derivatives of }\mathcal{P}\text{ are }\\
\text{discontinuous as consequence of }\\
\text{the ensemble inequivalence.}%
\end{array}
$} & \multicolumn{1}{|l|}{%
\begin{tabular}
[c]{c}%
1st-kind\\
dis$\text{continuous PT}$%
\end{tabular}
}\\\hline
\multicolumn{1}{|l|}{$%
\begin{array}
[c]{l}%
\text{Some of first derivatives of }\mathcal{P}\text{ are}\\
\text{discontinuous as consequence of }\\
\text{a discontinuity in some of the first }\\
\text{derivatives of the entropy }S\text{.}%
\end{array}
$} & \multicolumn{1}{|l|}{%
\begin{tabular}
[c]{c}%
2st-kind\\
dis$\text{continuous PT}$%
\end{tabular}
}\\\hline
\multicolumn{1}{|l|}{$%
\begin{array}
[c]{l}%
\text{Lost of analyticity of }\mathcal{P}\text{ with con-}\\
\text{tinuous first derivatives without }\\
\text{a lost of analyticity of entropy }S\text{.}%
\end{array}
$} &
\begin{tabular}
[c]{c}%
1st-kind\\
$\text{continuous PT}$%
\end{tabular}
\\\hline
\multicolumn{1}{|l|}{$%
\begin{array}
[c]{l}%
\text{Lost of analyticity of }\mathcal{P}\text{ with con-}\\
\text{tinuous first derivatives associated }\\
\text{to a lost of analyticity of entropy }S\text{.}%
\end{array}
$} &
\begin{tabular}
[c]{c}%
2st-kind\\
$\text{continuous PT}$%
\end{tabular}
\\\hline
\end{tabular}
\caption{Tentative classification scheme of the phase transitions (PT)
within a "canonical description" (open system).  \label{openned}}%
\end{table}%

Finally, the Table \ref{gen} summarizes the "genealogy" of the phase
transitions associated to the mathematical anomalies type \textbf{A} and
\textbf{B} of the microcanonical entropy $S$\ which were described in the
present section. While the anomaly type \textbf{A} (non concavity of $S$ in a
given reparametrization $\Theta$) is irrelevant when the interest system is
isolated, the incidence of certain experimental setup turns this behavior in a
lost of analyticity of the relevant thermodynamical potential $\mathcal{P}$
associated with the ensemble inequivalence (a 1st-kind discontinuous PT or its
limiting case: 1st--kind continuous phase transition). Type \textbf{B}
anomalies always leads to the lost of analyticity of $\mathcal{P}$ whenever
they are not hidden by the ensemble inequivalence. The corresponding phase
transitions are always originated from the occurrence of ergodicity breaking.
While microcanonical continuous phase transitions are directly related to the
2st-kind continuous phase transitions, a microcanonical discontinuous phase
transitions can manifest as a zero-order phase transition or a 2st-kind
discontinuous phase transition.
\begin{table}[tbp] \centering
\begin{tabular}
[c]{|c|c|c|}\hline\hline
\textbf{Anomaly} & \textbf{Isolated system} & \textbf{Open system}%
\\\hline\hline
\multicolumn{1}{|l|}{Type \textbf{A}} &
\begin{tabular}
[c]{c}%
microcanonically\\
irrelevant
\end{tabular}
& \multicolumn{1}{|l|}{%
\begin{tabular}
[c]{c}%
1st-kind discont. PT\\
1st-kind cont. PT
\end{tabular}
}\\\hline
\multicolumn{1}{|l|}{Type \textbf{B.I}} & micro. continuous PT & 2st-kind
cont. PT\\\hline
\multicolumn{1}{|l|}{Type \textbf{B.II}} & micro. discontinuous PT &
\begin{tabular}
[c]{c}%
zero-orden PT\\
2st-kind discont. PT
\end{tabular}
\\\hline\hline
\end{tabular}
\caption{Genealogy between the thermodynamic anomalies type A an B
under different external conditions. \label{gen}}%
\end{table}%

\section{Conclusions}

We have shown in the section \ref{RI}\ that the microcanonical description is
characterized by the presence of an internal symmetry whose existence is
closely related to the dynamical origin of this ensemble: the
\textit{reparametrization invariance}. Such symmetry leads naturally to a new
geometric formalism of the Thermodynamics within the microcanonical ensemble
which is not based on the consideration of a Riemannian metric derived from
the Hessian of the microcanonical entropy like other thermodynamic formalisms
proposed in the past \cite{rupper}. \ While the microcanonical entropy becomes
a scalar function within the present geometrical framework, we show that its
convex or concave character depends on the reparametrization of the integrals
of motion which are relevant for the microcanonical description. Such
ambiguity leads necessarily to a reconsideration of any classification scheme
of the phase transitions based on the concavity of the microcanonical entropy
\cite{gro1}.

Sections \ref{some}-\ref{scheme} were devoted to carry out a tentative
analysis of the above question. Interestingly, this aim demands a necessary
and unexpected generalization of the Gibbs canonical ensemble and the
classical fluctuation theory where the reparametrization invariance introduced
in the section \ref{RI}\ plays a more fundamental role than in the
conventional Thermodynamics. While reparametrization\ changes do not alter the
microcanonical description, such transformations represent specific
substitutions of the external experimental setup leading to different
generalized canonical descriptions \ref{g.canonical} for the open system.
Particularly, we have shown that the well-known ensemble inequivalence of the
Gibbs canonical ensemble \ref{gibbs} can be avoided within an appropriate
generalized canonical ensemble \ref{g.canonical}, which describes an open
system put in contact with a generalized version of the Gibbs thermostat (heat
bath) whose inverse temperature $\beta$ exhibits \textit{correlated
fluctuations} with the fluctuations of the total energy $E$ of the interest
system (see in subsection \ref{gen.can.ens}). This feature of the generalized
canonical ensemble put the basis for an unexpected generalization of the
classical fluctuation theory \cite{rupper} where the inverse temperature of
the generalized thermostat $\beta$\ and the total energy of the interest
system $E$ behave as \textit{complementary thermodynamic quantities} and the
convex regions derived from the entropy Hessian admit a simple interpretation:
there the inverse temperature $\beta$ can not be fixed since $\left\langle
\delta\beta\delta E\right\rangle \geq1$ (see in subsection \ref{unc.rel}). The
analysis of such questions demands a further study. Particularly, we have only
consider in the present work a generalization of the Gibbs canonical ensemble,
while it is reasonable a generalization of any Boltzmann-Gibbs distributions.

The possibility of avoid the ensemble inequivalence by using the generalized
canonical ensemble \ref{g.canonical} allows us to improve some Monte Carlo
methods based on the consideration of the Gibbs canonical ensemble. This aim
was carried out in the subsection \ref{gmmc} and the ref.\cite{vel-mmc} as a
example of application of the present reparametrization invariance ideas for
the implementation of a generalized version of the well-known Metropolis
importance sample algorithm \cite{met}.

The most important conclusion obtained from our subsequent analysis is the
recognition about the fundamental role of the arbitrary external conditions
(the using of different experimental setups) which can be imposed to the
interest system in order to control its thermodynamic equilibrium.
Particularly, the existence and nature of the phase transitions (admitting
them as a sudden change of the thermodynamic behavior of a given system while
a smooth change of a \textit{control parameter}) depends crucially on the
nature of the external experimental setup. This fact was shown in the section
\ref{potts} by considering the microcanonical description of the $q=10$ states
Potts model system, where the Gibbs canonical description predicts the
occurrence of a first-order phase transition and the microcanonical
description exhibits two microcanonical phase transitions which are hidden by
the ensemble inequivalence.

While there are only two thermodynamical anomalies which can be classified as
phase transitions within the microcanonical ensemble (presumably associated to
the occurrence of ergodicity breaking phenomenon \cite{Gold}), there are five
of them in an open system ("within a generalized canonical description"): two
of them closely related with the ensemble inequivalence (1st-kind
discontinuous PT and\ its limiting case, the 1st-kind continuous PT) and the
other three with the lost of analyticity of the entropy $S$ in the
thermodynamic limit (zero-order PT, 2st-kind discontinuous PT and the 2st-kind
continuous PT). To our viewpoint the present classification scheme is only
tentative and also deserves a further analysis of some paradigmatic models as
well as the revision of the available literature.

\appendix

\section{Riemannian metric from the Hessian of entropy\label{ruppeiner}}

\begin{figure}
[ptb]
\begin{center}
\includegraphics[
height=2.3774in,
width=2.1084in
]%
{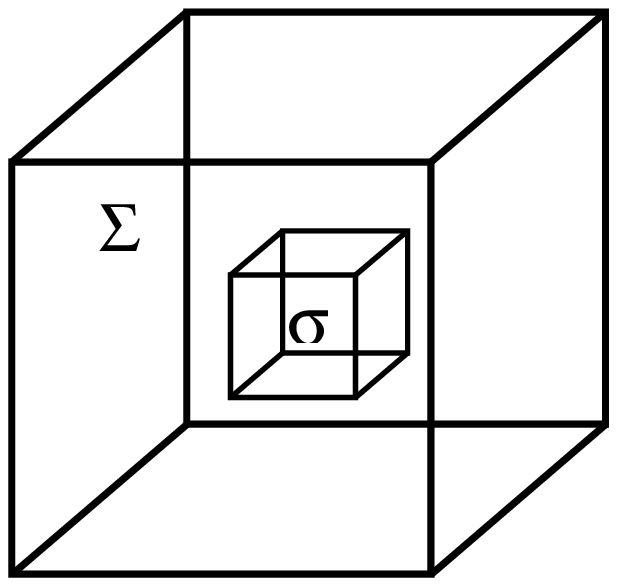}%
\caption{Usual setup in classical thermodynamics fluctuation theory and the
canonical ensembles in Statistical Mechanics: a subsystem $\sigma$ of a larger
closed enviroment $\Sigma$, which are ordinarily regarded as uniform systems.
}%
\label{diagram.eps}%
\end{center}
\end{figure}

Ruppeiner in ref.\cite{rupper} considers a subsystem $\sigma$ of a larger
closed system $\Sigma$ in a thermodynamic equilibrium characterized\ by the
macroscopic parameters $x_{0}$ ( a system + its environment - a situation
schematically represented in FIG.\ref{diagram.eps}). The probability that the
subsystem $\sigma$ be in a the macrostate characterized by the parameters $x$,
whose values are enclosed in the range $\left(  x,x+dx\right)  $ is given by:%

\begin{equation}
P\left(  x\left\vert x_{0}\right.  \right)  dx=\frac{1}{W\left(  x_{0}\right)
}\Omega\left(  x,x_{0}\right)  dx,
\end{equation}
being $\Omega\left(  x,x_{0}\right)  $ the density of states at the
configuration, and $W\left(  x_{0}\right)  =\int\Omega\left(  x,x_{0}\right)
dx$, the total number of microstates of the closed system. Instead of using
the Boltzmann entropy associated to the closed system $\Sigma$, $S_{B}\left(
x_{0}\right)  =\ln W\left(  x_{0}\right)  $, this author deals with the
"entropy" $S_{0}\left(  x,x_{0}\right)  =\ln\Omega\left(  x,x_{0}\right)  .$%

The most probable macroscopic configuration $x_{m}$ of the subsystem $\sigma$
can be derived as a solution of the stationary conditions:
\begin{equation}
\frac{\partial S_{0}\left(  x_{m},x_{0}\right)  }{\partial x^{\mu}}=0,
\label{station_rupper}%
\end{equation}
which ensures the nonnegative definition of the matrix $g_{\mu\nu}$ defined
from the Hessian :
\begin{equation}
g_{\mu\nu}\left(  x_{0}\right)  =-\frac{1}{V}\frac{\partial^{2}S_{0}\left(
x_{m},x_{0}\right)  }{\partial x^{\mu}\partial x^{\nu}}. \label{metric_rupper}%
\end{equation}
The matrix \ref{metric_rupper} is invariant under "reparametrizations" of the
thermodynamic parameters $x$ of the subsystem $\sigma$, $y=y\left(  x\right)
$, since the second terms of the transformation rule:%
\begin{align}
\frac{\partial^{2}S_{0}\left(  x\left(  y\right)  ,x_{0}\right)  }{\partial
y^{\alpha}\partial y^{\beta}}  &  =\frac{\partial x^{\mu}}{\partial y^{\alpha
}}\frac{\partial x^{\nu}}{\partial y^{\beta}}\frac{\partial^{2}S_{0}}{\partial
x^{\mu}\partial x^{\nu}}+\frac{\partial^{2}x^{\mu}}{\partial y^{\alpha
}\partial y^{\beta}}\frac{\partial S_{0}}{\partial x^{\mu}},\nonumber\\
&  \Rightarrow g_{\alpha\beta}\left(  x_{0}\right)  =\frac{\partial x^{\mu}%
}{\partial y^{\alpha}}\frac{\partial x^{\nu}}{\partial y^{\beta}}g_{\mu\nu
}\left(  x_{0}\right)  \label{rupper_rule}%
\end{align}
vanish as a consequence of the stationary conditions \ref{station_rupper}.
This feature allows to take the matrix $g_{\mu\nu}$ as a Riemannian metric
which provides important information about the fluctuations around the most
probable macroscopic configurations of the subsystem $\sigma$. The reader can
compare the result \ref{rupper_rule} with the one expressed in the equation
\ref{hessian_tr} in order to convince himself that the entropy Hessian of an
isolated system is not a Riemannian tensor.

\section{Coarsed grained microcanonical entropy\label{ent}}

For practical purposes, the Boltzmann entropy $S_{B}=\ln W$ can be estimated
by the coarsed grained entropy $S_{\varepsilon}=\ln\left\{  \Omega\left(
E,N\right)  \delta\varepsilon_{0}\right\}  $ where $\delta\varepsilon_{0}$ is
a very small energy constant which defines certain coarsed grained partition
$\mathcal{P}_{\varepsilon}$\ of the phase space $\Gamma$. The entropy in other
reparametrization $\Theta\left(  E\right)  =N\varphi\left(  E/N\right)
$\ could be estimated by $S_{\varphi}\left(  \Theta,N\right)  =\ln\left\{
\Omega\left(  \Theta,N\right)  \delta\varphi_{0}\right\}  \,$, where the small
constant $\delta\varphi_{0}$ defines another coarsed grained partition
$\mathcal{P}_{\varphi}$\ of the phase space. Obviously, the entropy estimates
$S_{\varepsilon}$ and $S_{\varphi}$ are not equal since the partitions
$\mathcal{P}_{\varepsilon}$ and $\mathcal{P}_{\varphi}$ are essentially
different. Taking into consideration the transformation rule
\begin{equation}
\Omega\left(  \Theta,N\right)  =\left\vert \frac{\partial\Theta}{\partial
E}\right\vert ^{-1}\Omega\left(  E,N\right)  ,
\end{equation}
the absolute discrepancy $\delta S$:
\begin{equation}
\delta S=S_{\varphi}\left(  \Theta,N\right)  -S_{\varepsilon}\left(
E,N\right)  =-\ln\left\{  \frac{\delta\varepsilon_{0}}{\delta\varphi_{0}}%
\frac{\partial\varphi\left(  \varepsilon\right)  }{\partial\varepsilon
}\right\}
\end{equation}
depends on the relative energy $\varepsilon=E/N$, but the relative discrepancy:%

\begin{equation}
\frac{\delta S}{S}\sim\frac{1}{N}\text{ with }S=\frac{1}{2}\left[
S_{\varepsilon}+S_{\varphi}\right]  ,
\end{equation}
drops to zero in the thermodynamic limit $N\rightarrow\infty$ as consequence
of the extensive character of the entropy when $N$ is large enough.

The above reasonings allows us to consider the coarsed grained entropy, or
more exactly, the entropy per particle in the thermodynamic limit:%
\[
s\left(  \varepsilon\right)  =\lim_{N\rightarrow\infty}\frac{S\left(
\varepsilon N,N\right)  }{N}=\lim_{N\rightarrow\infty}\frac{S\left(
\Theta\left(  \varepsilon N\right)  ,N\right)  }{N}=s\left(  \varphi\right)
\]
as a continuous scalar function whose form is independent from the nature of
the coarsed grained partitions $\mathcal{P}_{\varepsilon}$ and $\mathcal{P}%
_{\varphi}$ whenever their respective volume constants $\delta\varepsilon_{0}$
and $\delta\varphi_{0}$ be small enough. Thus, we can assume along the present
work the approximation $S\left(  E,N\right)  \cong S\left(  \Theta,N\right)  $.

\section{Others covariant quantities derived from the Boltzmann entropy
\label{others}}

Let us suppose that the integrals of motion depend on certain external
parameter $a$, $\hat{I}_{a}=\hat{I}\left(  X;a\right)  $. By taking the
partial derivative of the relation:
\begin{equation}
\Omega\left\langle \hat{O}\right\rangle =\int dX~\hat{O}\left(  X\right)
\delta\left[  I-\hat{I}\left(  X;a\right)  \right]  ,
\end{equation}
we obtain the identity:%
\begin{equation}
\frac{1}{\Omega}\frac{\partial}{\partial a}\left\{  \Omega\left\langle \hat
{O}\right\rangle \right\}  =\frac{1}{\Omega}\frac{\partial}{\partial I^{S}%
}\left\{  \Omega\left\langle \hat{O}\hat{F}_{a}^{S}\right\rangle \right\}  ,
\label{identity}%
\end{equation}
where the microscopic quantity $\hat{F}_{a}^{S}=\hat{F}_{a}^{S}\left(
X;a\right)  $ represents the \textit{k-th} component of the generalized force
associated to the parameter $a$:%
\begin{equation}
\hat{F}_{a}^{S}=-\frac{\partial\hat{I}_{a}^{S}}{\partial a},
\end{equation}

It is very easy to verify that the expectation values $\upsilon^{S}%
=\left\langle \hat{O}\hat{F}_{a}^{S}\right\rangle $ correspond to the
components of a contravariant vector because of they obey to the
transformation rule:%
\begin{equation}
\upsilon^{\prime K}=\frac{\partial\varphi^{K}}{\partial I^{S}}\upsilon^{S}.
\end{equation}
The differential operator:%
\begin{equation}
\mathcal{D}\left(  \mathbf{u}\right)  =\frac{1}{\Omega}\frac{\partial
}{\partial I^{S}}\left\{  \Omega u^{S}\right\}  , \label{div}%
\end{equation}
is just a linear transformation becoming a contravariant vector field
$\mathbf{u}=\left\{  u^{S}\right\}  $ in a scalar function.

The invariant volume element $d\mu_{g}$\ in the Riemannian geometry is derived
from the metric $g_{\mu\nu}$ as follows: $d\mu_{g}=\sqrt{g}d^{n}x$, where
$\sqrt{g}\equiv\sqrt{\left\vert \det g_{\mu\nu}\right\vert }$. The divergence
of a given vectorial field $\mathbf{u}$ in this framework is given by:
\begin{equation}
\operatorname*{div}\mathbf{u}=\frac{1}{\sqrt{g}}\frac{\partial}{\partial
x^{\mu}}\left\{  \sqrt{g}u^{\mu}\right\}  .
\end{equation}
The analogy between the invariant volume elements $d\mu_{g}=\sqrt{g}d^{n}x\sim
d\mu=\Omega dI$ allows us to consider the differential operator $\mathcal{D}%
\left(  \mathbf{u}\right)  $ \ref{div} as the divergence of a contravariant
vector field. Thus, it is obvious that the thermodynamic identity
\ref{identity} is reparametrization invariant.

Substituting $\hat{O}\equiv1$ and $\hat{O}\equiv\hat{F}_{a}^{T}$ in the
identity \ref{identity} and denoting by $S=\ln W$ where $W=\Omega\delta I$,
being $\delta I$ a small constant volume, the following equations are
straightforwardly obtained:
\begin{align}
\frac{\partial S}{\partial a}  &  =\frac{\partial S}{\partial I^{S}}F_{a}%
^{S}+\frac{\partial F_{a}^{S}}{\partial I^{S}},\label{eq1}\\
\frac{\partial S}{\partial a}F_{a}^{T}+\frac{\partial F_{a}^{T}}{\partial a}
&  =\frac{\partial S}{\partial I^{S}}F_{a}^{TS}+\frac{\partial}{\partial
I^{S}}F_{a}^{TS},
\end{align}
where we use the following nomenclature for the microcanonical averages
$F_{a}^{S}=\left\langle \hat{F}_{a}^{S}\right\rangle $ and $F_{a}^{TS}%
\equiv\left\langle \hat{F}_{a}^{T}\hat{F}_{a}^{S}\right\rangle $. The second
equation can be rephrased by introducing the correlation tensor $G_{a}%
^{TS}=F_{a}^{TS}-F_{a}^{T}F_{a}^{S}$ as follows:
\begin{equation}
\frac{\partial F_{a}^{T}}{\partial a}=\frac{\partial S}{\partial I^{S}}%
G_{a}^{TS}+\frac{\partial}{\partial I^{S}}G_{a}^{TS}+F_{a}^{S}\frac{\partial
F^{T}}{\partial I^{S}}. \label{eq2}%
\end{equation}
In the thermodynamic limit the quantities $\partial F_{a}^{S}/\partial I^{S}$
and $\partial G_{a}^{TS}/\partial I^{S}$ can be disregarded, and the quantity
$S$ becomes in the microcanonical entropy obtained from the coarsed grained of
the phase space. Thus, the thermodynamic identities:%
\begin{equation}
\frac{\partial S}{\partial a}=\frac{\partial S}{\partial I^{S}}F_{a}%
^{S},~\frac{\partial F_{a}^{T}}{\partial a}=\frac{\partial S}{\partial I^{S}%
}G_{a}^{TS}+F_{a}^{S}\frac{\partial F^{T}}{\partial I^{S}},
\end{equation}
express the relation among the entropy, the generalized forces, \textit{the
response functions} $\chi_{a}^{S}=\partial F_{a}^{S}/\partial a$ and the
corresponding correlation tensor $G_{a}^{TS}$ associated to certain external
parameter $a$ within the microcanonical ensemble. Particular expressions of
these results are the following:
\begin{align}
\frac{\partial S}{\partial V}  &  =\beta p,~\frac{\partial p}{\partial
V}=\beta G_{p}+p\frac{\partial p}{\partial E},\\
\frac{\partial S}{\partial B}  &  =\beta M,~\frac{\partial M}{\partial
B}=\beta G_{M}+M\frac{\partial M}{\partial E},
\end{align}
for a fluid ($p$ - pressure, $V$ - volume, $G_{p}$ - pressure dispersion) and
a magnetic system ($M$ - magnetization, $B$ - magnetic field, $G_{M}$ -
magnetization dispersion) respectively, where $\beta=\partial S/\partial E$ is
the microcanonical inverse temperature.


\begin{thebibliography}{99}                                                                                               %


\bibitem {gallavotti}G. Gallavotti, \textit{Statistical Mechanics}, (Springer,
Berlin, 1999).

\bibitem {Gold}N. Goldenfeld, \textit{Lectures on phase transitions and
critical phenomena}, Frontiers in physics 85, (Perseus Books Publishing,
L.L.C., 1992).

\bibitem {yang-lee}C.N. Yang and T.D. Lee, Phys. Rev. \textbf{87} (1952) 404.

\bibitem {moretto}L. G. Moretto \textit{et al}, Phys. Rep. \textbf{287} (1997) 249.

\bibitem {Dagostino}M. D'Agostino \textit{et al}, Phys. Lett. B \textbf{473}
(2000) 219.

\bibitem {gro1}D.H.E Gross, \textit{Microcanonical thermodynamics: Phase
transitions in Small systems}, \textit{66 Lectures Notes in Physics}, (World
scientific, Singapore, 2001).

\bibitem {gro na}D.H.E Gross and M. E. Madjet, Z. Physic B \textbf{104} (1997) 521.

\bibitem {PF1}H. Poincare, \textit{Les methodes nouvelles de la mecanique
celeste}, (Gauthier-Villars, Paris, 1892).

\bibitem {PF2}E. Fermi, Nuovo Cimento \textbf{25} (1923) 267; ibidem
\textbf{26} (1923) 105.

\bibitem {lieberman}A. J. Lichtenberg and M. A. Lieberman, \textit{Regular and
Stochastic Motion} (Springer, New York, 1993).

\bibitem {pettini 51}M. Cerruti-Sola and M. Pettini, Phys. Rev. E \textbf{51}
(1995) 53; Phys. Rev. E \textbf{53} (1996) 179.

\bibitem {cohenG}G. Gallavotti, E.G.D. Cohen, Phys. Rev. Lett. \textbf{74}
(1995) 2694 ; J. Stat. Phys. \textbf{80} (1995) 931.

\bibitem {rupper}G. Ruppeiner, Rev. Mod. Phys. \textbf{67} (1995) 605.

\bibitem {mc1}M. H. Kalos and P. A. Whitlock, \textit{Monte Carlo Methods} Vol
I: Basics (John Wiley \& Sons , 1986).

\bibitem {berg1}B. A. Berg, J. Stat. Phys. \textbf{82} (1996) 323.

\bibitem {berg2}B. A. Berg, Fields Inst. Commun. \textbf{26} (2000) 1.

\bibitem {met}N. Metropolis, A. W. Rosenbluth, M. N. Rosenbluth, A. H. Teller
and E. Teller, J. Chem. Phys. \textbf{21} (1953) 1087.

\bibitem {wang2}J. S. Wang, \textit{Efficient Monte Carlo Simulations Methods
in Statistical Physics}, e-print (2006) [cond-mat/0103318].

\bibitem {wolf}U. Wolff, Phys. Rev. Lett. \textbf{62} (1989) 361.

\bibitem {vel-mmc}L. Velazquez and J. C. Castro Palacio,\ \textit{Metropolis
Monte Carlo algorithm based on reparametrization invariance}, e-print (2006) [cond-mat/0606727].

\bibitem {pottsm}J. S. Wang, R. H. Swendsen and R. Koteck\'{y}, Phys. Rev.
Lett. \textbf{63} (1989) 1009; Phys. Rev. B \textbf{45} (1992) 4700.

\bibitem {gore}V. K. Gore and M. R. Jerrum, Proceeding of the 29th Anual ACM
Symposium on Theory of Computing (1997) 674; J. Stat. Phys. \ \textbf{97}
(1999) 67.

\bibitem {chava}P.H. Chavanis, Phys. Rev. E \textbf{65} (2002) 056123.
\end{thebibliography}
\end{document}